\documentclass[12pt,english]{article}
\usepackage[english]{babel}
\usepackage{amsmath,amssymb,amscd,color}
\usepackage{amsfonts}
\usepackage{graphicx}
\usepackage{url}
\usepackage{hyperref}
\usepackage{cite}
\usepackage{physics}

\makeatletter
\renewcommand\section{\@startsection {section}{1}{\z@}%
                                   {-5.5ex \@plus -1ex \@minus -.2ex}
                                   {2.3ex \@plus.2ex}%
                                   {\normalfont\large\bfseries}}
\renewcommand\subsection{\@startsection{subsection}{2}{\z@}%
                                     {-3.25ex\@plus -1ex \@minus -.2ex}%
                                     {1.5ex \@plus .2ex}%
                                     {\normalfont\bfseries}}

\numberwithin{equation}{section}

\makeatother

\newcommand{\bea}{\begin{eqnarray}}
\newcommand{\eea}{\end{eqnarray}}
\newcommand{\be}{\begin{equation}}
\newcommand{\ee}{\end{equation}}

\newcommand{\hf}{\frac{1}{2}}

\newcommand{\Z}{{\mathbb Z}}

\newcommand{\C}{{\mathbb C}}

\def\eg{{\it e.g.~}}

\newcommand{\cB}{{\cal B }}
            
\newcommand{\cM}{{\cal M }}            
\newcommand{\cF}{{\cal F }}

\newcommand{\cO}{{\cal O }}            
            
\newcommand{\cA}{{\cal A }}            
\newcommand{\cG}{{\cal G }}
\newcommand{\cN}{{\cal N }}

\newcommand{\cS}{{\cal S }}

\newcommand{\ie}{{\it i.e.~}}
\newcommand{\etc}{{\it etc.~}}

\newcommand{\mb}{{\bar m}}

\newcommand{\zb}{{\bar z}}
\newcommand{\xb}{{\bar x}}
\newcommand{\hb}{{\bar h}}
\newcommand{\qb}{{\bar q}}

\newcommand{\Jsf}{{\textrm J}}

\renewcommand{\title}[1]{\vbox{\center\LARGE{#1}}\vspace{5mm}}
\renewcommand{\author}[1]{\vbox{\center#1}\vspace{5mm}}
\newcommand{\address}[1]{\vbox{\center\footnotesize\em#1}}
\newcommand{\email}[1]{\vbox{\center\footnotesize\tt#1}\vspace{5mm}}

\begin{document}

\begin{titlepage}

 \begin{flushright}

\end{flushright}

\begin{center}

\hfill \\
\hfill \\
\vskip 1cm

\title{Threshold Corrections in Heterotic String Theory from Conformal Perturbation Theory
}

\author{Dilan Nur Demirta\c{s}$^a$, Christoph A. Keller$^b$, Manki Kim$^c$}

\author

\address{

${}^a$ Department of Physics, University of Arizona, Tucson, AZ 85721-0089, USA
\\~\\
${}^b$ Department of Mathematics, University of Arizona, Tucson, AZ 85721-0089, USA
\\~\\
${}^c$ Leinweber Institute for Theoretical Physics, Stanford University, Stanford, CA 94305, USA
}

\email{ dilandemirtas@arizona.edu, cakeller@arizona.edu, visionclavis@gmail.com}

\end{center}

\vfill

\abstract{We use conformal perturbation theory to compute one-loop threshold corrections to the K\"ahler potential of the heterotic string. For this we consider the theory compactified at the Gepner point of the quintic, that is at a highly stringy point of its moduli space. To obtain the corrections, we compute how the weights of the lightest states of the worldsheet theory shift at second order in conformal perturbation theory. This approach can be used to obtain the one-loop correction to all orders in $\alpha'$.}

\vfill

\end{titlepage}

\eject

\tableofcontents
\newpage

\section{Introduction}

Supersymmetric conformal field theories in 2d have a long history. They arise for instance as non-linear sigma models of Calabi-Yau manifolds. These are central ingredients for string compactifications, in which case the non-linear sigma model is a part of the worldsheet CFT that describes the dynamics of the string. For type II compactifications, the Calabi-Yau sigma model has $N=(2,2)$ supersymmetry \cite{Greene:1996cy}. CFTs with this much supersymmetry have another special property: they often come with a moduli space, that is they can be continuously deformed into other CFTs \cite{Lerche:1989uy}. Geometrically, this corresponds to deforming the Calabi-Yau manifold, either by changing its shape or complex structure, or its size or K\"ahler structure. There are thus K\"ahler moduli and complex structure moduli. From a CFT perspective, these deformations correspond to perturbing by exactly marginal fields, that is by fields of weight $(1,1)$ whose weight is protected from quantum corrections.  An unprotected field will usually  pick up such corrections to the weight so that it is no longer marginal. This introduces a renormalization group flow that breaks conformal invariance.
The reason why $N=(2,2)$ supersymmetry is special is that it is enough to protect against such quantum corrections.

The flip side to having a moduli space is that it is usually very hard to describe the CFT at an arbitrary point of the moduli space explicitly. One region of the moduli space that can be described is the large volume limit. Here the size of the Calabi-Yau manifold is large compared to the string length. It is thus well approximated by a free theory, possibly including some $\alpha'$ correction terms. Finding a description of the CFT at a stringy point in the moduli space, that is far away from the large volume limit, is much harder. Only in a few cases is a description known. One such class are the so-called Gepner models \cite{Gepner:1987qi,Gepner:1989gr}. These are theories that are composed of $N=(2,2)$ minimal models. They are rational and therefore in principle solvable. However, there are only a few such Gepner models. To explore the vast remainder of the moduli space, a  different approach is needed.

One such approach is conformal perturbation theory. As described above, from a CFT perspective, moving around the moduli space corresponds to perturbing by exactly marginal fields. Perturbation theory allows to compute order by order in the perturbation parameter the changes of the CFT under this deformation, for instance the shift of its spectrum. In practice this allows to at least explore the neighborhood of a given point in the moduli space by computing the first few terms of the perturbative expansion. 

One motivation for doing this is to investigate the question of how common rational CFTs are in the moduli space. There is a conjecture due to \cite{Gukov:2002nw} that rational CFTs are dense in the moduli space of K3, but not for higher dimensional Calabi-Yau manifolds. For toroidal theories, that is CFTs that are essentially free theories, this can be addressed relatively directly \cite{Moore:1998pn,Wendland:2000ye,Kidambi:2022wvh,Okada:2022jnq}, but for non-toroidal  Calabi-Yau manifolds, other tools are needed. One possible approach is to use conformal perturbation theory to check where new conserved currents might appear, which could render the CFT rational \cite{Benjamin:2020flm}. For K3, this idea was investigated in  \cite{Keller:2023ssv}. In view of the above conjecture, it makes sense to repeat this analysis for a  Calabi-Yau threefold, to compare and contrast to the K3 case. In this article we indeed set up the conformal perturbation theory formalism for the quintic, which would allow such an analysis. However, we will pursue a different question here.

The driving motivation of this article is a different one: we want to investigate compactifications of the heterotic string on the quintic. In general, computing physical observables of the compactification and in particular their stringy corrections is difficult. This has only been done for quantities that are protected by supersymmetry such as in \cite{Bershadsky:1993cx,Antoniadis:1993ze,Antoniadis:1996qg,Antoniadis:1996vw,Antoniadis:1997eg}, or for toroidal models such as in \cite{Berg:2004ek,Berg:2005ja,Haack:2015pbv}. The issue is that we can only do such computation if we know the sigma model CFT explicitly. As pointed out above, away from the large volume limit we generally do not know it. Among the few exceptions to this are Gepner point, where the  CFTs are known explicitly and described by a rational theories.
There we can thus at least in principle compute physical quantities such as the K\"ahler potential and its corrections.

Our goal is to compute the string one-loop corrected K\"ahler metric of moduli fields in heterotic string theory compactified on a Calabi-Yau threefold at a Gepner point. One motivation is that to obtain any cosmologically reliable constructions in string theory, it may be necessary to stabilize moduli; this in turn  by definition requires fixing vacuum expectation value of moduli fields at non-trivial value, which then necessitates the understanding of perturbative control \cite{Dine:1985he}. 
Although it is harder to achieve moduli stabilization in heterotic string theory \cite{Gukov:2003cy}, Calabi-Yau compactifications in heterotic string theory provide a fruitful arena to study the $g_s$ corrections to the low-energy effective action. First, unlike type II string theory compactified on Calabi-Yau threefolds, heterotic compactifications only enjoy $\mathcal{N}=1$ spacetime supersymmetry, so that the K\"ahler metric cannot be computed exactly. Second, unlike type II string theories on Calabi-Yau orientifolds that also enjoy $\cN=1$ supersymmetry, in heterotic string theory we are only concerned with closed strings. This simplifies detailed studies. Therefore, heterotic string theory sits in the sweet spot where the first loop corrections to the kinetic action of the moduli are not protected, yet relatively simple to compute.

In this paper, we shall study the one-loop correction to the kinetic action of moduli fields through the threshold corrections to the gauge coupling. To do this, we will follow the approach of \cite{Kaplunovsky:1995jw}. 
Although it is easy to state the recipe of this one-loop calculation, the list of explicit one-loop calculation is limited despite its importance. For example, the only explicit one-loop corrections known to the best of the authors' knowledge were carried out in toroidal compactifications. For more general Calabi-Yau manifolds, this calculation is notoriously difficult even in the large volume limit, as one does not know the explicit metric, and much less the string spectrum --- see \cite{Anguelova:2010ed} for an example of such a large volume computation from a spacetime perspective.
Here we want to perform such a computation at a highly stringy point: the Gepner point of the quintic threefold. Although the one-loop effects are expected to be stringy, we are still able to carry out the calculations, as the worldsheet theory is known.

Let us be more concrete and also explain how this relates to conformal perturbation theory. The idea is to
compute the threshold correction $\Delta_a$  \cite{Kaplunovsky:1995jw}, 
\begin{align}\label{Deltaaintro}
    \Delta_a :=\int \frac{d^2\tau}{\tau_2} (B_a-b_a)\,.
\end{align}
Here
\begin{align}\label{eqn:Bintro}
    B_a:= -\frac{1}{\eta(\bar\tau)^2} \sum_{even~ s}\text{Tr}_{(\tilde{R},s_1)} \left((-1)^{s_2\bar F+F}F q^{L_0-\frac{3}{8}}\bar{q}^{\bar{L}_0-\frac{3}{8}}\right)_{int}Z(s_1,s_2,G_a)\, ,
\end{align}
where the trace is over the spectrum of the internal CFT, which depends on the moduli; $b_a$ is a moduli-independent constant.
The correction $\Delta_a$ can then be related to the K\"ahler potential through Green-Schwarz anomaly type arguments.
We are interested in the K\"ahler metric, that is the second derivative of the K\"ahler potential. This can be related to the second derivative of $\Delta_a$, see for instance (2.15) in \cite{Kaplunovsky:1995jw}. 
This second derivative in turn can be obtained by using second order perturbation theory to compute the the shift of the internal spectrum, allowing us to obtain the K\"ahler metric.

In this paper we explain how to perform the computation of this second order shift. For convenience, we work with the $N=(2,2)$ CY compactification of the quintic at the Gepner point, as the shift of this spectrum can be easily related to the heterotic spectrum. We then compute the lifting matrix of the lightest non-BPS states. To do this, we need to evaluate certain Gepner model four-point functions and then integrate them. Gepner models are composed of $N=2$ minimal models, whose correlation functions we thus need to calculate. On a technical level, doing this for the quintic is harder than for the quartic discussed in \cite{Keller:2023ssv}: For the quartic, the minimal models have central charge $c=3/2$ and can be related to a free boson and an Ising model (or, equivalently, a free fermion). The correlation functions are then relatively elementary functions, at most involving square roots. For the quintic, the equivalent description involves parafermions. Alternatively, minimal models can be described by $SU(2)$ WZW models and free bosons. How to obtain these WZW model correlation functions was described in \cite{Zamolodchikov:1986bd}. We work out explicit expressions for the needed correlation functions, which in our case turn out to be hypergeometric functions.

We then pick a modulus $\Phi$ and deform the CFT by $\mu \Phi$.
To obtain the shift in the spectrum, we match the perturbed two-point function in the usual way to the general form, 
\be
\langle\varphi^\dagger (z_1)\varphi(z_2)\rangle_\mu = \langle\varphi^\dagger(z_1)\varphi(z_2) e^{\mu\int_\epsilon d^2w \Phi(w) + c.t.}\rangle\ .
\ee
Evaluating the perturbed two-point function order by order in its $\mu$-expansion leads to integrated $n$-point functions. These integrals are divergent when two fields collide and therefore need to be regulated. In standard conformal perturbation theory this is for instance done by cutting out $\epsilon$-discs and adding counterterms to the Lagrangian to cancel the divergences; this is what the notation $\int_\epsilon$ signifies. To obtain the lifting matrix we can then perform the integrals as in \cite{Keller:2023ssv}. After dealing with the divergences in this way, we are essentially left with just the constant term of the integral.

This brings us to a second aspect of this article.
Even though this regularization scheme works and gives the correct answer at second order, it is somewhat naive and ad hoc. In particular, generalizing it to higher order would be more subtle and would introduce technical difficulties, such as dealing with the case when multiple fields collide. It would therefore be nice to have more principled scheme.
It was pointed out in \cite{Sen:2019jpm} that string field theory can provide such a scheme. In particular, string field theory does not suffer from any divergences. This means that if we perform this computation carefully in the framework of string field theory, in particular keeping track of picture changing operators (PCO), the result should be finite. This implies that string field theory implicitly gives a regularization scheme for conformal perturbation theory consistent to all orders. In the process of our calculations, we check explicitly that at second order, this string field theory scheme agrees with the naive scheme we introduced above. This follows the spirit of other recent work that has investigated conformal perturbation theory from the perspective of string field theory \cite{Scheinpflug:2023osi,Mazel:2024alu}.

Let us briefly summarize the outcome. In principle, our approach allows us to compute the $\alpha'$-exact one-loop correction in $g_s$ to the K\"ahler metric. For practical reasons though, we achieve something slightly less here: 
To compute the full one-loop correction to the K\"ahler metric as in (\ref{eqn:torus two2}), we would have to compute the shift of all the states in the spectrum of the theory. We will be less ambitious here and instead only compute the shift of the lightest non-protected states, which turn out to have $h=(1/5,1/5)$. We will also not evaluate the integral (\ref{eqn:torus two2}). Even though the lightest states whose lifting we compute will presumably make the most important contribution to the integral, heavier states may make significant contributions as well, so that the integral of just the lightest states would not give reliable information. We therefore content ourselves with setting up the formalism and giving a proof of principle for our approach.75, leaving heavier states and evaluating the integral to future work.

The outline of this paper is as follows: In section~\ref{s:threshold}, we set up our conventions and give the explicit expression for the threshold corrections in terms of the $\cN=(2,2)$ CY sigma model spectrum. We also explain how the K\"ahler metric is related to the shift of the states at second order perturbation theory. In section~\ref{s:confpert} we first set up conformal perturbation theory with a conventional regularization scheme. We then show that this scheme agrees with the string field theory computation. In section~\ref{s:Gepner} we introduce Gepner models and discuss the spectrum of the quintic at the Gepner point. In section~\ref{s:corr} we explain how to compute the corresponding correlation functions. In section~\ref{s:shift} we then evaluate the integral and compute the shift of the lightest states in the spectrum. Appendix~\ref{app:stringvertices} explains our choice of local coordinates for computing string field theory correlators, and appendix~\ref{app:corr} gives some background on computing four-point functions in the $SU(2)$ WZW models that are needed for our computations.

\section{Threshold corrections and the K\"ahler potential}\label{s:threshold}
\subsection{Strategy}
Let us explain the strategy we will take to compute the one-loop corrected K\"ahler metric of the moduli space in heterotic string theory at the Gepner point, for which the loop corrections are expected to be the most sizable.

In principle string field theory provides a systematic prescription to compute such corrections directly.  Although this approach may become fruitful, one-loop computations in string field theory are still difficult to perform. Therefore, we instead take a different route. For heterotic string theory, one-loop corrected K\"ahler potential is related to an appropriate one-loop partition function via Green-Schwarz anomaly type arguments \cite{Kaplunovsky:1995jw}. We can therefore compute the one-loop corrected K\"ahler metric using (somewhat schematically) the relation
\begin{equation}\label{eqn:torus two2}
    K_{\mu\bar{\mu}}= \partial_\mu\partial_{\bar{\mu}} Z\ ,
\end{equation}
where $Z$ is the partition function. We will give more details on this below.
To first order in the $g_s^2$ expansion, the string spectrum that enters \eqref{eqn:torus two2} is independent of $g_s$ corrections. We therefore conclude that if we can compute the partition function $Z$ and at the same time compute the shift of the spectrum in $\mu$ and $\bar{\mu},$ we can compute the one-loop corrected K\"ahler metric without going through the direct string field theory evaluation.

\subsection{Worldsheet conventions}
Let us collect some useful formulas and conventions for the worldsheet theory and string field theory. 
We shall study heterotic string theory compactified on a Calabi-Yau threefold with the standard embedding. For the intermediate steps of the calculations, we shall use type IIB string theory to compute the spectrum in a deformed background. For type IIB string theory, the anti-holomorphic sector is simply the complex conjugate of the holomorphic sector. The worldsheet theory of the heterotic string theory consists of a $(\mathcal{N},\overline{\mathcal{N}})=(1,0)$ matter CFT with central charge $(15,26),$ a $b,c,\beta,\gamma$ ghost system for the holomorphic sector, and a $\bar{b},\bar{c}$ ghost system for the anti-holomorphic sector. We shall take the matter CFT to be a direct sum of a $(\mathcal{N},\overline{\mathcal{N}})=(1,0)$ free CFT of central charge $(6,4)$ for the four non-compact directions, a $(\mathcal{N},\overline{\mathcal{N}})=(2,2)$ Calabi-Yau CFT with central charge $(9,9)$, and an anti-holomorphic free fermion system of central charge $(0,13)$ 

The $(\mathcal{N},\overline{\mathcal{N}})=(1,0)$ free matter CFT with central charge $(6,4)$ consists of the matter fields $X$ and $\psi,$ in the NS sector with OPE 
\begin{equation}
    X^\mu (z) X^\nu(0)\sim -\frac{\alpha'}{2} \eta^{\mu\nu}\log|z|^2\,,\quad \psi^\mu(z)\psi^\nu(0) \sim \frac{\eta^{\mu\nu}}{z}\,.
\end{equation}
The free fermion CFT of central charge $(0,13)$ consists of 26 worldsheet fermions $\bar{\lambda}^a$ with OPE
\begin{equation}
    \bar{\lambda}^a(\bar{z})\bar{\lambda}^b(0)\sim \frac{\delta^{ab}}{z}\,.
\end{equation}
The $\beta,~\gamma$ ghost system we bosonize by introducing the scalar field $\phi$ and the fermionic fields $\xi,~\eta$ as
\begin{equation}
    \beta=\partial \xi e^{-\phi}\,,\quad \gamma=\eta e^\phi\,.
\end{equation}
The OPE of the ghost system is
\begin{align}
    &c(z)b(0)\sim \frac{1}{z}\,,\quad \xi(z)\eta(0) \sim\frac{1}{z}\,,\\
    &\partial\phi(z)\partial\phi(0)\sim-\frac{1}{z^2}\,,\quad e^{q_1\phi(z)}e^{q_2\phi(0)}\sim z^{-q_1q_2}e^{(q_1+q_2)\phi(0)}\,.
\end{align}
Combining these ingredients, we introduce the total worldsheet stress energy tensor $T$ and the total worldsheet supercurrent $G$. For the holomorphic side they decompose as
\begin{align}
    &T=T_m+T_{\text{gh}}=T_m^f+T_m^{\text{int}}+T_{\text{gh}}\,,
    & G=G_m+G_{\text{gh}}=G_m^f+G_m^{\text{int}}+G_{\text{gh}}\,,
\end{align}
where we split up the contribution into a matter part and a ghost part. The matter part in turn we split into a contribution of the holomorphic part of the 4d free field CFT,
\begin{equation}
T_m^f=-\frac{1}{\alpha'}\partial X^\mu\partial X_\mu-\frac{1}{2}\psi_\mu\partial \psi^\mu\,,
\end{equation}
\begin{equation}
    G_m^f=i\sqrt{\frac{2}{\alpha'}{2}} \psi^\mu \partial X_\mu\,,
\end{equation}
and a contribution of the $c=9$, $\cN=2$ Calabi-Yau CFT $T^{\text{int}}_m$ and $G^{int}_m$.
In particular, as the internal CFT enjoys extended supersymmetry, we have two internal supercurrents $G_+$, $G_-$, and an $U(1)_R$ current $J$. 
The internal part of the supercurrent is given as a linear combination of $G_+$ and $G_-$
\begin{equation}\label{eqn:Gint norm}
    G_{m}^{\text{int}}=\frac{1}{\sqrt{2}} (G_++G_-)\,.
\end{equation}
The internal superconformal generators have the OPEs
\begin{align}\label{intOPEfirst}
T_m^{\text{int}}(z) T_m^{\text{int}}(w)=&\frac{9}{2(z-w)^4}+\frac{2T_m^{\text{int}}(w)}{(z-w)^2}+\frac{\partial T_m^{\text{int}}(w)}{z-w}+\dots\,,\\
J(z)J(w)=&\frac{3}{(z-w)^2}+\dots\,,\\
J(z)G_+(w)=&\frac{1}{z-w}G_+(w)+\dots\,\\
J(z)G_-(w)=&-\frac{1}{z-w}G_-(w)+\dots\,\\
G_+(z) G_-(w)=&\frac{6}{(z-w)^3}+\frac{2J(w)}{(z-w)^2}+\frac{1}{z-w}(\partial J(w)+2T_m^{\text{int}}(w))+\dots\,\\
G_+(z)G_+(w)=&\text{regular}\,,\\
G_-(z)G_-(w)=&\text{regular}\,. \label{intOPElast}
\end{align}
Moreover we can bosonize the $U(1)$ R current 
\begin{equation}
J=i\sqrt{3}\partial H\,,
\end{equation}
such that a vertex operator $V$ with charge $q$ can be written as
\begin{equation}
V= e^{i\frac{q}{\sqrt{3}}H} \tilde{V}\,,
\end{equation} 
where $\tilde{V}$ is neutral. Note that we have
\begin{equation}
L_0 e^{i\frac{q}{\sqrt{3} }H}=-\frac{q}{12}e^{i\frac{q}{\sqrt{3} }H}\,.
\end{equation}
In the anti-holomorphic sector we have
\be
\bar T=\bar T_m^f+\bar T_m^{\text{int}}+ \bar T_g+ \bar T_{\text{gh}}\ ,
\ee
where $\bar T_g$ is the contribution of the free gauge fermion system
\be
\bar T_g = -\frac12 \bar\lambda_a \bar\partial \bar \lambda^a\ ,
\ee
$\bar T_{\text{gh}}$ is the contribution of the $\bar b, \bar c$ ghost system and
\be
\bar T_m^f = -\frac{1}{\alpha'}\bar\partial X^\mu\bar \partial X_\mu\ .
\ee
There is no overall anti-holomorphic supercurrent, but the internal CFT has (super-) currents $\bar G_+, \bar G_-, \bar J$ satisfying the OPEs (\ref{intOPEfirst}) through (\ref{intOPElast}).

We write the BRST currents as
\begin{align}\label{jBRST}
    &j_B=c\left( T_m-\frac{1}{2}(\partial\phi)^2-\partial^2\phi-\eta\partial \xi\right) +\eta e^\phi G_m+bc\partial c-\eta\partial\eta be^{2\phi}\,,\\
    &\bar{j}_B =\bar{c}\bar{T}_m+\bar{b}\bar{c}\bar{\partial}\bar{c}\,,
\end{align}
and the BRST charge as
\begin{equation}
    Q_B=\frac{1}{2\pi i}\oint j_B-\frac{1}{2\pi i}\oint \bar{j}_B\,.
\end{equation}
Next, we define the picture changing operator (PCO) as 
\begin{equation}\label{PCO}
    \mathcal{X}:=\{Q_B,\xi\}=c\partial\xi +e^\phi G_m-\partial \eta be^{2\phi}-\partial(\eta be^{2\phi})\,.
\end{equation}
Now we shall determine the GSO projection to decide which operators are admissible in our analysis. We shall define the holomorphic worldsheet fermion number $f$ that counts the number of worldsheet fermions in the NS sector. We shall define $f$ such that chiral spin fields in the Ramond sector have $f=1/2$ mod $2.$ Also, we assign the fermion number $q$ to the operator $e^{q\phi}$.
We define the anti-holomorphic worldsheet fermion number $\bar{f}$ that counts the number of $\bar{\lambda}.$ We then define the GSO parity operator $F$ and $\bar{F}$ 
\begin{equation}
    F:= f+Q\,,\quad \bar{F} := \bar{f}+\bar{Q}\,,
\end{equation}
where $Q$ and $\bar{Q}$ count the $U(1)_R$ charge of the internal sector. We then define the GSO projection such that only the operators that are even under $(-1)^F$ and $(-1)^{\bar{F}}$ are allowed in the spectrum. 

An important remark on the GSO projection we chose is in order. The spectrum of the Gepner model can be obtained as follows. First, one constructs a type II worldsheet theory of a Calabi-Yau compactification whose GSO projection is given by the direct analogue of the GSO parity operator we constructed. Then, the one-loop partition function of such a worldsheet CFT can be shown to be modular invariant. The idea of Gepner \cite{Gepner:1987qi} was then to use the fact that the modular transformation properties of the partition function of the $(\mathcal{N},\overline{\mathcal{N}})=(0,1)$ free field CFT with central charge $(0,6)$ and the $\bar{b},~\bar{c},~\bar{\beta},~\bar{\gamma}$ ghost system is the same as that of the direct sum of a free field fermionic CFT with central charge $(0,(24+2)/2),$ four anti-chiral bosons, and the $b,~c$ ghost system. Naturally, after replacing the partition function of the former CFT with the latter without changing the spin sum in the anti-holomorphic sector, the resulting partition function as a whole is again modular invariant. One can check that the GSO projection implicit in the construction of Gepner's partition function is the same as the GSO projection we chose. 

Next, let us explain how to find the marginal operators of heterotic string theory starting from a type II construction. In type II string compactifications, marginal deformation operators describing Calabi-Yau moduli in the $(-1,-1)$ picture are written as
\begin{equation}
    V^{-1,-1}=c\bar{c} e^{-\phi}e^{-\bar{\phi}} V\,,
\end{equation}
where $V$ is an operator of the internal matter CFT with $(h,\bar{h})=(1/2,1/2)$ and $(Q,\bar{Q})=(\pm 1,\pm1).$ 
The operator of the heterotic string theory in the $-1$ picture can be recovered by collecting the bosonic pieces of the $(-1,0)$ picture operator of type II. For example, we can act with an anti-holomorphic PCO
\begin{equation}
    \overline{\mathcal{X}}_0 V^{-1,-1} \supset -\frac{1}{2\pi i} \oint d\bar{z}  c\bar{c} e^{-\phi} G_m V= c\bar{c} e^{-\phi} Y\,,
\end{equation}
where we defined
\begin{equation}
    Y:=-\frac{1}{2\pi i}\oint d\bar{z} \overline{G}_m V\,.
\end{equation}
The operator
\begin{equation}
    c\bar{c}e^{-\phi}Y\,,
\end{equation}
is then a GSO even operator that corresponds to the marginal deformation in heterotic string theory.

\subsection{Computing the one-loop corrected K\"ahler metric in the heterotic string}
In this section, we want to write down the character formula for the one-loop corrected K\"ahler metric in heterotic string compactification with the standard embedding. 
The goal is to compute the one-loop corrected K\"ahler metric of moduli fields in a heterotic string compactification on a Calabi-Yau threefold with the standard embedding at the Gepner point. As we chose the standard embedding for the vector bundle, the internal part of the CFT is that of a $(\cN,\bar\cN)=(2,2)$ Calabi-Yau CFT with central charge 9. To compute the one-loop corrected K\"ahler metric, we need the second-order variation of the spectrum around the Gepner point. As the only spectrum that is affected by the moduli deformation is captured by the Calabi-Yau CFT, for the calculation of the shifted spectrum, we use perturbation theory for a type II string compactification on the same Calabi-Yau manifold. 

Our overall strategy for computing the corrections is based on \cite{Kaplunovsky:1995jw}. The idea is to relate the derivative of the K\"ahler potential to the derivative of the string-threshold corrections $\Delta_a$. These in turn are determined by the spectrum of the compactification. More concretely, the one-loop partition function we shall evaluate, at the shifted vacuum, is given by 
\begin{align}\label{Deltaa}
    \Delta_a :=\int \frac{d^2\tau}{\tau_2} (B_a-b_a)\,,
\end{align}
where
\begin{align}\label{eqn:B}
    B_a:= -\frac{1}{\eta(\bar\tau)^2} \sum_{even~ s}\text{Tr}_{(\tilde{R},s_1)} \left((-1)^{s_2\bar F+F}F q^{L_0-\frac{3}{8}}\bar{q}^{\bar{L}_0-\frac{3}{8}}\right)_{int}Z(s_1,s_2,G_a)\,.
\end{align}
Here the trace contains the contribution of the internal states of the Calabi-Yau CFT, and $Z(s_1,s_2,G_a)$ is the contribution of the gauge group,
\begin{equation}
    Z(s_1,s_2,G_a):=\text{Tr}_{s_1}\left((-1)^{s_2\bar F}\left(T_a^2-\frac{1}{8\pi\tau_2}\right)\bar q^{\bar L_0-\frac{13}{24}}\right)\,.
\end{equation}
The sum runs over all spin structures $(s_1,s_2)$ on the anti-holomorphic side except for $(1,1)$. On the holomorphic side the only spin structure that contributes is $\tilde R = (1,1)$. $b_a$ is an invariant of the gauge group, which is independent of the moduli. We will therefore not further discuss it.

Next, let us study which states of the Calabi-Yau CFT contribute to the one-loop amplitude. The GSO projection for the holomorphic sector is given by
\begin{equation}
    (-1)^{F+F_{gh}+F_{f}}\,,
\end{equation}
where $F_{gh}$ and $F_{f}$ are the fermion number operators for the ghost sector and the free field CFT, respectively, and $F$ is the $U(1)_L$ charge of the internal CFT. Note that the vacuum state
\begin{equation}
    ce^{-\phi}
\end{equation}
is GSO odd. Hence, states with even fermion number of the CY CFT must be paired with the odd number of free fermions to create GSO even states. As is evident from the formula, every holomorphic state in the Ramond sector contributes regardless of its GSO parity. However, one should not interpret this as a statement that GSO odd spacetime fields can run in the loop. Rather, such states should be understood as the KK excitations of gravitons, for example. 

On the other hand, the GSO projection for the anti-holomorphic sector is given by
\begin{equation}
    (-1)^{\bar F+\bar F_{g}}\,,
\end{equation}
where $\bar F$ is the $U(1)_R$ charge of the internal CFT and $\bar F_g$ is the fermion number of the free-field fermions for the unbroken gauge group. The formula \eqref{eqn:B} involves the sum over the spin structure for the anti-holomorphic sector. As the GSO projection involves not only the $U(1)_R$ charge of the internal CFT, but also the unbroken gauge factor, we again conclude that any state of the internal CFT can contribute to the loop. There is one crucial difference between the states with even $\bar F$ and odd $\bar F.$ Suppose that there is a state $\mathcal{O}$ with trivial $U(1)_R$ charge and weight $\bar h.$ This state can directly run in the loop, contributing $\bar q^{\bar h-3/8},$ as $\mathcal{O}$ alone is GSO even. However, if there is a state $\mathcal{O}$ with a non-trivial $U(1)_R$ charge, this state must be paired with a fermion of the unbroken gauge factor, thereby increasing the effective weight by $1/2.$   

Now we shall organize the partition function in terms of the characters of the extended $\mathcal{N}=2$ superconformal algebra.
As mentioned before, the gauge partition function is independent of the moduli and is in fact given by
\begin{equation}
    Z(s_1,s_2,E_6)=\frac{\vartheta_{00}(\bar\tau)^8+\vartheta_{10}(\bar\tau)^8+\vartheta_{01}(\bar\tau)^8}{\eta(\bar\tau)^8}  \frac{\vartheta_{s_1s_2}(\bar\tau)^5}{\eta(\bar\tau)^5}\times\frac{1}{2\pi i} \frac{\partial}{\partial\bar \tau}\left( \log \frac{\vartheta_{s_1s_2}(\bar\tau)}{\eta(\bar\tau)}+\frac{1}{2} \log(\tau_2|\eta(\tau)|^4)\right)\,.
\end{equation}
Next, we want to write the trace over the internal states in terms of characters of the of the extended $\mathcal{N}=2$ superconformal algebra. As our goal is to study the change of the partition function under an exactly marginal deformation, we shall only include the massive representations.
The character formula for a massive irreducible representation with weight $h$ and the charge $Q$ in the $(s_1,s_2)$ spin structure is given by \cite{Odake:1988bh,Odake:1989dm,Odake:1989ev}
\begin{equation}
    g_{s_1s_2}^{(h,Q)}=(-1)^{s_2|Q|} \frac{q^{h-\frac{1+|Q|}{4}}}{\eta(\tau)^3} \vartheta_{s_1s_2}(\tau) \vartheta_{s_1+|Q|,0}(2\tau)\,.
\end{equation}
Note that the $U(1)_R$ charge can take the value $0,~1,~-1.$ Further note that here we define $h$ and $Q$ the weight and the $U(1)_R$ charge of the primary state in the NS sector. This gives us the contribution from the holomorphic sector. To obtain the contribution of the holomorphic sector to (\ref{eqn:B}), we pick the $(1,1)$ structure and take a derivative with respect to fermion number fugacity to pull down a fermion number $\bar F$, giving \cite{Kim:2023cbh}
\be\label{Zh}
  Z_{h,Q_L}=   (-1)^{|Q_L|+1} q^{h-\frac{1+|Q_L|}{4}} \vartheta_{1+|Q_L|,0}(2\tau)\,.
\ee
Gathering the sum over the spin structure for the holomorphic part we obtain
\begin{align}\label{Zhbar}
   \bar Z_{\bar h,Q_R} =&\sum_{even\ s} g_{s_1s_2}^{(\hb,Q_R)}(\bar\tau)
    Z(s_1,s_2,E_6)\,,
\end{align}
so that
\be
B_a= -\frac1{\eta(\bar\tau)^2} \sum_{h,\hb,Q_L,Q_R} Z_{h,Q_L}(\tau) \bar Z_{\hb,Q_R}(\bar \tau)\ .
\ee
We now want to compute $\partial_\mu \partial_{\bar\mu} B_a$. The only dependence on the moduli $\mu,\bar\mu$ is in the internal weights $h,\hb$ of the massive primary fields.
To compute the second derivative, we have 
\be
\partial_\mu q^{h}\qb^{\hb} = (2\pi i \tau \partial h -2\pi i \bar \tau \partial \hb) q^h \qb^{\hb} = -4\pi \Im(\tau) \partial h q^h\qb^\hb
\ee
where we used that $\partial h = \partial \hb$, since the spin of a state is integer and remains unchanged under perturbations, and
\be\label{Kaehlercontribution}
\partial_\mu \partial_{\bar \mu} q^{h}\qb^{\hb} = -4\pi \Im(\tau) \bar \partial \partial h q^h\qb^\hb 
+ 16\pi^2 \Im(\tau)^2 \partial h \bar \partial h q^h \qb^\hb
\ee
We see that in order to compute the corrections to the K\"ahler metric, we need to obtain the first and second order derivatives of the internal weights of states. These can be obtained by conformal perturbation theory at first and second order. We now to turn to this. We will see that the first order contributions actually vanish, so that only the genuine second order term contributes.

\section{Conformal perturbation and string field theory}\label{s:confpert}
\subsection{Conformal perturbation theory}\label{ss:naiveCPertT}
Let us now discuss how to use conformal perturbation theory to compute the shift in the spectrum. We will first give a quick overview of `conventional' conformal perturbation theory  \cite{Cardy:1987vr,Dijkgraaf:1987jt,Kutasov:1988xb}, that is, perturbation theory of CFTs that do not necessarily come from string field theory. As is the case with perturbation theory, we need to introduce a regulator $\epsilon$, a regularization scheme and appropriate counterterms. We will choose hard sphere regularization. From a purely CFT point of view, this choice is somewhat ad hoc, as we could have made a different choice here. The belief is however that different schemes give the same physical results and only give a different choice of coordinates on the moduli space. However, in later sections we will show that string field theory actually gives a natural regularization scheme \cite{Sen:2019jpm}, which we check agrees with the naive scheme here up to the order we need. 

We are interested in the shift of the conformal weight of primary fields under perturbation by an exactly marginal field $\Phi$, the modulus. For more details, see \cite{Benjamin:2020flm} and \cite{Keller:2023ssv}.
For this we schematically expand the two-point function 
\be\label{2ptpert}
\langle\varphi^\dagger (z_1)\varphi(z_2)\rangle_\mu = \langle\varphi^\dagger(z_1)\varphi(z_2) e^{\mu\int d^2w \Phi(w)}\rangle
= \frac{1}{(z_1-z_2)^{2h(\mu)}(\zb_1-\zb_2)^{2\hb(\mu)}}
\ee
in powers of the coupling $\mu$.  From the correlator (\ref{2ptpert}) we can read off the shift in the conformal dimensions $(h,\hb)$ of $\varphi$.  To be more precise, we are looking to compute the coefficients $h^{(n)}$ in the expansion
\be\label{hexpand}
h(\mu)= \sum_{n=0}^\infty h^{(n)} \mu^n\ ,
\ee
where $h^{(0)}$ is the dimension of the field in the unperturbed theory. Since the spin is integral and therefore constant under perturbations, the exact same expression (except for the $\hb^{(0)}$ term) must hold for $\hb(\mu)$. 

To obtain the expressions for the $h^{(n)}$, we expand (\ref{2ptpert}) on both sides and match the terms. This implies essentially that $h^{(n)}$ is given by an $n$-fold integral of a $n+2$-point function. However, using conformal symmetry, we can fix one integration variable, leaving a $n-1$-fold integral. For the first order term for instance this leads to the well-known result that
\be\label{h1shift}
h^{(1)} = -\pi C_{\varphi^\dagger\varphi\Phi}\ .
\ee
At second order, we find the integral
\be\label{h2integral}
h^{(2)} =-\frac\pi2 M^{(2)} = -\frac{\pi}{2} \int d^2 x \,\langle \varphi^\dagger(\infty)\Phi(1)\Phi(x)\varphi(0)\rangle\ .
\ee
Our discussion so far has been very schematic. In particular, we did not deal with the divergences when $\Phi(x)$ collides with the other fields and their regularization. A more precise form of (\ref{2ptpert}) is
\be\label{2ptpertreg}
\langle\varphi^\dagger (z_1)\varphi(z_2)\rangle_\mu = \langle\varphi^\dagger(z_1)\varphi(z_2) e^{\mu\int_\epsilon d^2w \Phi(w) + c.t.}\rangle
= \frac{A(\epsilon, \mu)}{(z_1-z_2)^{2h(\mu)}(\zb_1-\zb_2)^{2\hb(\mu)}}\ .
\ee
Here $\int_\epsilon$ indicates a regularized version of the integral with regularization parameter $\epsilon$ and $c.t.$ indicates some ($\epsilon$ dependent) counterterms. $A(\epsilon,\mu)$ finally is the wave function renormalization of $\varphi$.

We use a hard-sphere regularization scheme that cuts out $\epsilon$-discs around $0,1,\infty$. We then cancel the resulting $\epsilon$-divergences by subtracting counterterms (at second order)
\be\label{naiveCT}
-\sum_{\Delta_k<2} \mu^2 \frac{2\pi C_{\Phi\Phi\Psi_k}}{2-\Delta_k} \epsilon^{\Delta_k-2} \int d^2 w \Psi_k(w)
+ \sum_{\Delta_k=2} \mu^2 2\pi C_{\Phi\Phi\Psi_k} \log (\epsilon)  \int d^2 w \Psi_k(w)\ .
\ee
The second order term is thus given by the constant term of the integral (\ref{h2integral}), namely
\be
M^{(2)} = \int_{\C_\epsilon} d^2 x \,\langle \varphi^\dagger(\infty)\Phi(1)\Phi(x)\varphi(0)\rangle\ ,
\ee
where $\C_\epsilon$ is the complex plane with discs cut out,
\be\label{Cepsilon}
\C_{\epsilon}= \{ z \in \C : |z|>\epsilon, |z-1|>\epsilon, |z|< \epsilon^{-1}\} .
\ee

\subsection{String field theory}
Let us now discuss an alternative approach to regularization. String field theory provides a natural worldsheet UV regulator. This is an attractive choice as string field theory guarantees that the worldsheet UV regulator used is self-consistent. We will check that at least up to second order, this regulator provides the same prescription as in section~\ref{ss:naiveCPertT}. 

The remainder of section~\ref{s:confpert} is not strictly necessary for the rest of the article and can be skipped on first reading. We will therefore be fairly brief, especially in introducing our notation. For a review on string field theory, see for example \cite{deLacroix:2017lif,Sen:2024nfd}; for notation and conventions, see \cite{Sen:2014dqa,Sen:2015hha}.

To construct heterotic string field theory, we first define the state space. We denote by $H_{p}$ a space of GSO even string states with picture number $p.$ Then we define $H_T$ and $\tilde{H}_T$ as
\begin{equation}
    H_T:=H_{-1}\oplus H_{-1/2}\,,
\end{equation}
\begin{equation}
    \tilde{H}_T:=H_{-1}\oplus H_{-3/2}\,.
\end{equation}
By $\Psi$ and $\tilde{\Psi}$ we denote states in $H_T$ and $\tilde{H}_T$ respectively. Note that the half integer picture states denote Ramond states. We define an operator $\mathcal{G}$ that acts on states in $\tilde{H}_T$ 
\begin{equation}
    \mathcal{G} \tilde{\Psi}=\tilde{\Psi}\,,
\end{equation}
for $\tilde{\Psi}\in H_{-1}$ and
\begin{equation}
    \mathcal{G}\tilde{\Psi}= \mathcal{X}_0 \tilde{\Psi}\,,
\end{equation}
for $\tilde{\Psi}\in H_{-3/2}$, where $\mathcal{X}_0$ is the zero mode of the picture changing operator (\ref{PCO}).

Next, we define string vertices $\{\}$. Following \cite{Sen:2014dqa,Sen:2015hha}, we define the 1PI region of the moduli space to exclude the region of the moduli space that cannot be represented by 1PI Feynman diagrams of string field theory. We declare that the three-point correlator, which has no modulus, is 1PI. Also, we declare that the entire moduli space of torus one-point function is 1PI. Following \cite{Sen:2015hha}, we first define string vertices $\{\Psi_0\Psi_1\ldots \Psi_N \}$, and then also  string brackets $[\cdots ]$ through 
\begin{equation}
    \langle \Psi_0|c_0^-|[\Psi_1\dots \Psi_N]\rangle:= \{ \Psi_0\Psi_1\dots \Psi_N\}\, .
\end{equation}
Using this notation, the 1PI off-shell action of heterotic string field theory is given by
\begin{equation}
    S= g_s^{-2}\left( -\frac12 \langle \tilde{\Psi}|c_0^-Q_B|\tilde{\Psi}\rangle +\langle \Psi|c_0^-Q_B\mathcal{G}|\tilde{\Psi}\rangle +\sum_n \frac{1}{n!} \left\{\Psi^n\right\}\right) \,.
\end{equation}
This gives the equation of motion for $\Psi$ as
\be
Q_B |\Psi\rangle + \sum_{n=1}^\infty\frac{1}{(n-1)!} \cG [ \Psi^{n-1}]= 0\ .
\ee
We will now use this to study vacuum shifts and quantum amplitudes.

\subsection{Background shift}\label{ss:backgroundshift}
In this section, we study the deformation by marginal operators in the context of string field theory. The goal is to study the background field equations of motion perturbatively to second order. 

We will treat $\mu$ and $\bar{\mu}$ as small expansion parameters, and we expand the background solution $\Psi^0, \tilde \Psi^0$ as
\begin{equation}
    \Psi^0=\sum_{n,m} \mu^n\bar{\mu}^m \Psi_{n,m}^0\,,
\end{equation}
\begin{equation}
    \tilde{\Psi}^0=\sum_{n,m}\mu^n\bar{\mu}^m\tilde{\Psi}_{n,m}^0\,.
\end{equation}
Here $\Psi^0_{0,0}=\tilde\Psi^0_{0,0}=0$ because they represent the original background.
Without loss of generality, we fix $\mathcal{G}\tilde{\Psi}^0=\Psi^0.$ Although for our discussion the distinction between type IIA and type IIB and also the distinction between (c,c) and (c,a) moduli is not meaningful, for concreteness we choose to work with type IIB string theory. For the background we study, the first order terms $(n,m)=(1,0)$ and $(0,1)$ are described by (c,c) and (a,a) operators of the chiral ring, respectively.
That is,
\begin{equation}\label{Vmodulus}
    \Psi^0_{1,0}= c\bar{c} e^{-\phi}e^{-\bar{\phi}} V\,,
\end{equation}
\begin{equation}
    \Psi^0_{0,1}=c\bar{c} e^{-\phi} e^{-\bar{\phi}} W\,,
\end{equation}
where $V$ and $W$ are vertex operators of dimension $(1/2,1/2)$,  and their $R$ charges are
\begin{equation}
    V(z,\bar{z}) W(y,\bar{y})=-\frac{1}{|z-y|^2}\,,
\end{equation}
\begin{equation}
    \oint dz J(z) V(0)=\oint d\bar{z}\bar{J}(\bar{z}) V(0)=V(0)\,,  
\end{equation}
\begin{equation}
    \oint dz J(z) W(0)=\oint d\bar{z} \bar{J}(\bar{z}) W(0)=-W(0)\,.
\end{equation}
Hence $V$ and $W$ are (c,c) and (a,a) operators with charge $\pm 1$ respectively. Note that we included a factor of $1/(2\pi i)$ in $\oint.$

Let us now study the background equation up to the order $(n,m)=(1,1).$ For our analysis, we will not need $(n,m)=(2,0)$ and $(0,2).$ 
The first-order background equations are given as
\begin{equation}
    Q_B |\Psi_{1,0}^0\rangle=Q_B|\Psi_{0,1}^0\rangle=0\,,
\end{equation}
which are automatically solved by definition. Here we use the fact that $\Psi^0_{0,0}=\tilde\Psi^0_{0,0}=0$ and that the one-point functions vanish. 
The second-order background equation is given as
\begin{equation}
    \frac{4}{g_c^2} Q_B |\Psi_{1,1}^0\rangle= - [ \Psi_{1,0}^0 \otimes \Psi_{0,1}^0]_{S^2}\,.
\end{equation}
Moreover
\begin{equation}
    \Bbb{P}\Psi_{1,1}^0=0\,,
\end{equation}
as we are deforming the background by exactly marginal deformations. 

\subsection{Computing the the type II string spectrum}
In this section, we shall lay out the recipe to compute the shift of the Calabi-Yau spectrum. As we illustrated before, we are after the change of the spectrum under $\partial_\mu\partial_{\bar{\mu}}.$ 
Hence, the goal is to find a solution to the linearized equation of motion in the shifted vacuum. That is, we take $\Psi=\Psi_0+\delta \Psi$, where we assume that $\Psi_0$ solves the equation of motion, and linearize the resulting equation of motion in $\delta \Psi$. This will allow us to find a solution for $\delta \Psi$ which in turn will allow us to compute the shift of the weight of some primary field $\cO$.

The computation of the partition function requires the shift of both the GSO even and GSO odd states of the internal CFT. Suppose that a conformal primary $\mathcal{O}$ of the internal CFT is GSO even, meaning, 
\begin{equation}
(-1)^{F}\mathcal{O}=(-1)^{\bar{F}}\mathcal{O}=-\mathcal{O}\,.
\end{equation}
We then want to solve the linearized equation of motion of the string field at the shifted background whose 0-th order form is written as
\begin{equation}
    \delta\Psi_{0,0}= c\bar{c}e^{-\phi}e^{-\bar{\phi}}\mathcal{O} e^{ik\cdot x}\,.
\end{equation}
For a GSO odd conformal primary $\mathcal{O}_o$ of the internal CFT, on the other hand, we shall solve the linearized equation of motion for the following seed state
\begin{equation}
    \delta\Psi_{0,0}=c\bar{c} e^{-\phi} \psi^\mu e^{-\bar{\phi}}\bar{\psi}^\nu \mathcal{O}_o e^{ik\cdot x}\,.
\end{equation}
In both cases, by computing the change of the mass-shell condition, we can read off the shift of the internal CFT spectrum.

We will now illustrate how to compute the spectrum shift of $\mathcal{O}.$  First we expand the shift in the background in $\mu$ and $\bar\mu$ as above,
\begin{equation}
    \delta\Psi= \sum_{n,m}\mu^n\bar{\mu}^m\delta\Psi_{n,m}\, .
\end{equation}
We then solve the linearized equations of motion perturbatively in $\mu$ and $\bar{\mu}$. The relevant linearized equations now read
\begin{equation}\label{order0}
    Q_B| \delta\Psi_{0,0}\rangle=0\,,
\end{equation}
\begin{equation}\label{order1}
    Q_B |\delta\Psi_{1,0}\rangle=-[\Psi_{1,0}^0 \otimes \delta\Psi_{0,0}]_{S^2}\,, \quad Q_B|\delta\Psi_{0,1}\rangle=-[\Psi_{0,1}^0\otimes\delta\Psi_{0,0}]_{S^2}\,,
\end{equation}
\begin{align}\label{order2}
   Q_B| \delta\Psi_{1,1}\rangle=&-[\Psi_{1,1}^0\otimes\delta\Psi_{0,0}]_{S^2}- [\Psi_{1,0}^0\otimes\Psi_{0,1}^0\otimes \delta\Psi_{0,0}]_{S^2}\nonumber\\
    &-[\Psi_{1,0}^0\otimes \delta\Psi_{0,1}]_{S^2}-[\Psi_{0,1}^0\otimes \delta\Psi_{1,0}]_{S^2}\,.
\end{align}
These are the equations up to order 2. We will now discuss how to solve them to obtain the shift in the spectrum.

\subsubsection{Zeroth order}
Using the form of the BRST current (\ref{jBRST}), the zeroth order equation of motion (\ref{order0}) is equivalent to
\begin{equation}
    \frac{\alpha'}{4}k^2+(h-1/2)=0\,,
\end{equation}
where $h$ is the weight of $\mathcal{O}.$ We will use it to connect the momentum $k$ to the weight $h$. In particular we can use it to compute the shift in weight as
\be\label{deltah}
\delta h = - \frac{\alpha'}{2}k \delta k\ .
\ee

\subsubsection{First order}
Let us now study the first-order equations (\ref{order1}). We shall study $\delta\Psi_{1,0}$ in detail and write the results for $\delta\Psi_{0,1}.$ We shall first define a projection operator $\Bbb{P}$ that projects states into $L_0^+=0$ components and the rest. Then, we can write the first order equation as
\begin{equation}\label{eqn:first order eom}
    \Bbb{P}Q_B|\delta\Psi_{1,0}\rangle=-\Bbb{P}[\Psi_{1,0}^0\otimes\delta\Psi_{0,0}]_{S^2}\,,
\end{equation}
and
\begin{equation}
    (1-\Bbb{P})Q_B|\delta\Psi_{1,0}\rangle=-(1-\Bbb{P}) [\Psi_{1,0}^0\otimes\delta\Psi_{0,0}]_{S^2}\,.
\end{equation}
Because BRST operator is invertible for $L_0^+\neq0$ states
\begin{equation}
    \{b_0^+,Q_B\}=L_0^+\,,
\end{equation}
we can solve the $(1-\Bbb{P})$ component of the equation easily
\begin{equation}\label{eqn:first order 1-p sol}
    (1-\Bbb{P})|\delta\Psi_{1,0}\rangle= -\frac{b_0^+}{L_0^+}(1-\Bbb{P})[\Psi_{1,0}^0\otimes \delta\Psi_{0,0}]_{S^2}\,.
\end{equation}
The above equation implies that the marginal deformation forces mixing between the original state and arbitrary off-shell states that have a non-trivial overlap against the source term. Note, however, that below we will take the large stub limit, in which the mixing of the form \eqref{eqn:first order 1-p sol} in general has vanishing contributions. 

To calculate the source term on the RHS of \eqref{eqn:first order eom}, we compute an overlap between a test string field and the source term. Here, we only need to study the shift of the scalar spectrum, so that we pick the test field
\begin{equation}\label{testfield}
    \mathcal{V}_t:= c\bar{c} e^{-\phi}e^{-\bar{\phi}} \mathcal{O}_te^{-ik\cdot x}\,,
\end{equation}
where $\mathcal{O}_t$ is an operator of dimension $h$.  All the other level zero vertex operators, for example,
\begin{equation}
    c\bar{c} (\eta e^{-2\bar{\phi}}\bar{\partial}\bar{\xi}- e^{-2\phi}\partial\xi\bar{\eta})\,,
\end{equation}
have zero overlaps due to the ghost number anomaly. Note that the ghost-dilaton does not have a non-trivial overlap with the source term, so we shall ignore this in the first-order analysis.

The overlap with the source term is thus given by the  string vertex 
\begin{equation}
    \{ \Psi_{1,0}^0 \otimes \delta\Psi_{0,0}\otimes \mathcal{V}_t \}_{S^2}\,.
\end{equation}
To compute it, we use the fact that all of the string fields inserted in the above bracket are on-shell. Therefore, we can place the PCOs as we like, as their locations does not affect the result. We act with both the holomorphic and anti-holomorphic PCOs on $\Psi_{1,0}^0$,
\begin{equation}\label{eqn:00 pic first order}
    \mathcal{X}\overline{\mathcal{X}} \Psi_{1,0}^0\supset c\bar{c} V^{(1,1)}\,,
\end{equation}
where we define $V^{(1,1)}$ as
\begin{equation}
    V^{(1,1)}:= -\frac{1}{(2\pi i)^2} \oint dz G_m\oint d\bar{z} \overline{G}_m V\,.
\end{equation}
Here $V$ is the (c,c) modulus introduced in (\ref{Vmodulus}).
Then we compute
\begin{align}
    \{\Psi_{1,0}^0 \otimes \delta\Psi_{0,0}\otimes \mathcal{V}_t \}_{S^2}=& \left\langle c\bar{c}V^{(1,1)}(0)\otimes c\bar{c} e^{-\phi}e^{-\bar{\phi}} \mathcal{O}(1) \otimes c\bar{c} e^{-\phi}e^{-\bar{\phi}} \mathcal{O}_t(\infty) \right\rangle_{S^2}\,.
\end{align}
Note that we used the $SL(2;C)$ symmetry to fix the locations of the vertices. Note also that the additional terms in \eqref{eqn:00 pic first order} do not contribute due to the ghost number anomaly. As one can check, the above string vertex will give a number 
\begin{equation}
    \delta\mathcal{N}_{1,0}:=\{\Psi_{1,0}^0 \otimes \delta\Psi_{0,0}\otimes \mathcal{V}_t \}_{S^2}=-C_{V^{(1,1)}\mathcal{O}\mathcal{O}_t}\,.
\end{equation}
We can then write the corresponding string bracket as
\begin{equation}\label{sourceterm}
    -\Bbb{P}\left[\Psi_{1,0}^0\otimes \delta\Psi_{0,0}\right]_{S^2}= \delta\mathcal{N}_{1,0}  (\partial c+\bar{\partial}\bar{c})c\bar{c} e^{-\phi}e^{-\bar{\phi}} \mathcal{O}e^{ik\cdot X}\,.
\end{equation}
To solve the equation of motion, we use the ansatz
\begin{align}\label{delta10ansatz}
    \Bbb{P}\delta\Psi_{1,0}=& c\bar{c} e^{-\phi}e^{-\bar{\phi}} \mathcal{O} (e^{i(k+\delta k_{1,0})\cdot X}-e^{ik\cdot X})\,,
\end{align}
which gives 
\begin{equation}\label{QBacting}
    Q_B \Bbb{P}\delta\Psi_{1,0}=\frac{\alpha'}{2} k\cdot \delta k_{1,0}(\partial c+\bar{\partial}\bar{c}) c\bar{c} e^{-\phi}e^{-\bar{\phi}} \mathcal{O} e^{i(k+\delta k_{1,0})\cdot X}\,.
\end{equation}
Matching (\ref{QBacting}) and (\ref{sourceterm}) gives
\begin{equation}
    \frac{\alpha'}2 k\cdot \delta k_{1,0}=\delta \mathcal{N}_{1,0}=-C_{V^{(1,1)}\mathcal{O}\mathcal{O}_t}\,.
\end{equation}
By picking the test operator $\cO_t$ as $\cO^\dagger$ and using (\ref{deltah}), we obtain 
\be
\delta h =  C_{V^{(1,1)}\mathcal{O}\mathcal{O}^\dagger}\ .
\ee
This agrees with \eqref{h1shift} after taking into account the normalization of $V^{1,1)}$ compared to $\Phi$.

\subsubsection{Second order}
Let us now now repeat this for shift of the spectrum at second order perturbation. The equation of motion we need to solve is (\ref{order2}),
\begin{multline}\label{secondorderQB}
\Bbb{P} Q_B|\delta\Psi_{1,1}\rangle= \\
-\Bbb{P} \left[ \Psi_{1,1}^0\otimes \delta\Psi_{0,0} + \Psi_{1,0}^0\otimes \Psi_{0,1}^0\otimes \delta\Psi_{0,0}+ \Psi_{1,0}^0\otimes\delta\Psi_{0,1}+\Psi_{0,1}^0\otimes \delta\Psi_{1,0}\right]_{S^2}\,.
\end{multline}
In order to evaluate the resulting correlation functions, we need to introduce local coordinates. We give explicit expressions for those in appendix~\ref{app:stringvertices}. What is important here is that they contain a stub parameter $\lambda$. 
While the answer is independent of the choice of $\lambda$, to make our computations easier we will eventually take the large stub limit $\lambda\to\infty$. Because $\Bbb{P}\Psi_{1,1}^0$ is trivial, this will then in particular allow us to neglect the first term on the right-hand side. Of the other terms, let us first focus on the third term
\begin{equation}
    -\Bbb{P}[\Psi_{1,0}^0\otimes \delta\Psi_{0,1}]_{S^2}\,.
\end{equation}
To compute this, we shall again evaluate a vertex with a test field,
\begin{equation}
\mathcal{A}_2=    -\{ \Psi_{1,0}^0\otimes \delta\Psi_{0,1}\otimes \mathcal{V}_t\}_{S^2}\,.
\end{equation}
Now, as $\delta\Psi_{0,1}$ is not an on-shell operator, we need to carefully choose the location of PCO. Furthermore, $\mathcal{A}_2$ will depend on our choice of local coordinates. We will place the PCOs at the symmetric locations
\begin{equation}
    p_\pm=\frac{1\pm i\sqrt{3}}{2}\, ,
\end{equation}
and $\Psi_{1,0},$ $\delta\Psi_{0,1}$ and $\mathcal{V}_t$ at $0,~1,~\infty,$ respectively --- see appendix~\ref{app:stringvertices} for details.
The off-shell amplitude $\mathcal{A}_1$ is then determined to be 
\begin{align}
    \mathcal{A}_2=&- \biggr\langle c\bar{c} e^{-\phi}e^{-\bar{\phi}} V(0) \otimes \lambda^{-L_0-\bar{L}_0}c\bar{c} e^{-\phi} e^{-\bar{\phi}} \mathcal{O} (e^{i(k+\delta k_{0,1})\cdot X}-e^{ik\cdot X})(1)\nonumber\\&\qquad\qquad \otimes c\bar{c} e^{-\phi}e^{-\bar{\phi}} \mathcal{O}_t e^{-ik\cdot X}(\infty) \otimes \mathcal{X}(p_\pm) \otimes \overline{\mathcal{X}}(\bar{p}_\pm)\biggr\rangle_{S^2}\,.
\end{align}
Here we used (\ref{testfield}), (\ref{delta10ansatz}) and (\ref{Vmodulus}) as expressions for the three fields.
The stub parameter $\lambda$ parametrizes the local coordinates.
To evaluate the above amplitude, we note that the zero mode integral is non-trivial only when the second term of
\begin{equation}
\lambda^{-L_0-\bar{L}_0}=1-(L_0+\bar{L}_0)\log \lambda+\dots
\end{equation}
acts on factors such as $e^{i(k+\delta k_{0,1})\cdot X}$. Therefore we arrive at 
\begin{align}
    \mathcal{A}_2=&2\alpha'(k\cdot \delta k_{0,1}) \log\lambda \biggr\langle c\bar{c} e^{-\phi}e^{-\bar{\phi}} V(0) \otimes c\bar{c} e^{-\phi} e^{-\bar{\phi}} \mathcal{O} e^{i k\cdot X}(1)\nonumber\\&\qquad\qquad \otimes c\bar{c} e^{-\phi}e^{-\bar{\phi}} \mathcal{O}_t e^{-ik\cdot X}(\infty) \otimes \mathcal{X}(p_\pm) \otimes \overline{\mathcal{X}}(\bar{p}_\pm)\biggr\rangle_{S^2}\,.
\end{align}
As the remainder of the CFT correlators are all on-shell, we can freely move the PCOs. We shall bring them to $V(0)$ to arrive at
\begin{align}
    \label{eqn:(1,1) log}\mathcal{A}_2=&2\alpha'(k\cdot \delta k_{0,1})\log\lambda \left\langle c\bar{c} V^{(1,1)}(0)\otimes c\bar{c} e^{-\phi}e^{-\bar{\phi}} \mathcal{O} e^{ik\cdot X}\otimes c\bar{c} e^{-\phi}e^{-\bar{\phi}}\mathcal{O}_te^{-ik\cdot X}(\infty)\right\rangle_{S^2}\,,\\
    =&4 C_{V^{(1,1)}\mathcal{O}\mathcal{O}_t}C_{W^{(1,1)}\mathcal{O}\mathcal{O}_t} \log\lambda\,.
\end{align}
Similarly, there is a contribution $\cA_3$ coming from the last term in (\ref{secondorderQB}), which differs from the above by exchanging holomorphic and anti-holomorphic, that is $V$ by $W$\etc

Next, we calculate the contribution of the second term,
\begin{equation}
    -\Bbb{P}[\Psi_{1,0}^0\otimes\Psi_{0,1}^0\otimes \delta\Psi_{0,0}]_{S^2}\,.
\end{equation}
To calculate the above source term, we compute the overlap
\begin{equation}
    \mathcal{A}_1= -\{\mathcal{V}_t \otimes \Psi_{1,0}^0\otimes \Psi_{0,1}^0 \otimes\delta\Psi_{0,0}\}_{S^2}\,.
\end{equation}
To evaluate $\mathcal{A}_1,$ we insert two holomorphic PCOs and two anti-holomorphic PCOs. As all the operators that are inserted in $\mathcal{A}_1$ are on-shell, one may naively conclude that one can place PCOs anywhere in the four punctured Riemann surface. However, a generic choice of PCOs will not satisfy the boundary conditions for the PCO locations determined by the three-point functions. In particular, in the degeneration channel where the four punctured Riemann sphere degenerates into two three punctured Riemann spheres, we need to correctly place one PCO in each of the three punctured Riemann spheres. The error caused by the wrong PCO distributions in the degeneration limit can be fixed by performing vertical integrations of PCOs on the boundary of the moduli space \cite{Sen:2014pia, Sen:2015hia}. Therefore we will place one holomorphic and one anti-holomorphic PCOs at the background fields in the interior of the moduli space, and fix the error caused by this naive choice of PCOs by performing vertical integrations on the boundary of the moduli space.

In the interior of the moduli space we compute
\begin{equation}
 \cA_1 =\frac{1}{\pi}  \label{eqn:4pt}\int_{\mathcal{M}} d^2z \langle c\bar{c} e^{-\phi} e^{-\bar{\phi}}\mathcal{O} e^{ik\cdot x}(1)\otimes c\bar{c} e^{-\phi}e^{-\bar{\phi}}\mathcal{O}_te^{-ik\cdot x}(\infty) \otimes c\bar{c} V^{(1,1)}(0)\otimes W^{(1,1)}(z)\rangle \,, 
\end{equation}
where $\mathcal{M}$ is determined to be the region by the following inequalities
\begin{equation}\label{lambdacutouts}
   \left| \frac{16 e^{i\theta} \lambda^{-2}}{(4+e^{i\theta} \lambda^{-2})^2}\right| <|z|\,,\quad \left| \frac{16 e^{i\theta} \lambda^{-2}}{(4+e^{i\theta} \lambda^{-2})^2}\right|  <|z-1|\,,\quad \left| \frac{16 e^{i\theta} \lambda^{-2}}{(4+e^{i\theta} \lambda^{-2})^2}\right|  <|1/z|\,,
\end{equation}
where $\theta$ can take any real value --- see appendix~\ref{app:4ptsphere} for a more detailed description of this choice of local coordinates. The main point of this section is that this integral corresponds to the usual four-point function integral that appears at second order in conformal perturbation theory. The prescription (\ref{lambdacutouts}) corresponds to regularizing the integral by cutting out neighborhoods of roughly size $\lambda^{-2}$ around the divergent points $0,1,\infty$. 

String field theory guarantees that the total result is independent of the choice of parametrization of the local coordinates and hence independent of $\lambda$, once all other contributions are taken into account.
However, for the explicit evaluation of the above formula, it is useful to take the large $\lambda$ limit and compute terms order by order in $\lambda.$
This gives terms that are divergent in $\lambda$, a constant term, and terms that vanish in this limit.
The claim is that the divergent terms will all be cancelled if we perform our procedure carefully, giving a finite result and thus providing a good regularization scheme. Some of these cancellations come from the contributions of $\cA_2$ and $\cA_3$. The remaining divergences are cancelled by the contribution of vertical integration \cite{Sen:2014pia, Sen:2015hia}. 
In the end we will thus be left with the constant term in (\ref{eqn:4pt}), agreeing with the result in section~\ref{ss:naiveCPertT}.

To be more explicit about these divergences, we split $\cM$ into regions $\cM_{0,1,\infty}$ around the divergent points $0,1,\infty$. We will split off the diverging contributions $\mathcal{B}_0, \mathcal{B}_1, \mathcal{B}_\infty$ to $\cM_0,~\cM_1,~\cM_\infty$ which are generated around $|z|\sim \lambda^{-2}\,,~|z-1|\sim\lambda^{-2}\,,~ |z|\sim\lambda^2,$ respectively. 

Let us first study $\cM_0.$ The OPE of the marginal deformations $V^{(1,1)}$ and $W^{(1,1)}$ produces divergent terms
\begin{equation}
    W^{(1,1)}(z) V^{(1,1)}(0)\sim \sum_{\Delta_\varphi\leq 2} \frac{C_{W^{(1,1)}V^{(1,1)}\varphi}}{|z|^{4-\Delta_\varphi}}\varphi(0)\,,
\end{equation}
that lead to
\begin{equation}
    \mathcal{B}_0= - \frac{1}{\pi}\sum_{\Delta_\varphi\leq2}\int_{|z|\geq  \lambda^{-2}} d^2z \frac{C_{\mathcal{O} \mathcal{O}_t \varphi} C_{W^{(1,1)}V^{(1,1)}\varphi^c}}{|z|^{4-\Delta_\varphi}}\,. 
\end{equation}
Similarly, $\mathcal{B}_1$ and $\mathcal{B}_\infty$ can be written as
\begin{equation}
    \mathcal{B}_1=-\frac{1}{\pi}\sum_{\Delta_\varphi\leq \Delta_\mathcal{O}} \int_{|z-1|\geq \lambda^{-2}}d^2z \frac{C_{\mathcal{O}W^{(1,1)}\varphi } C_{\mathcal{O}_t V^{(1,1)}\varphi^c}}{|z-1|^{2+\Delta_\mathcal{O}-\Delta_\varphi}}\,,
\end{equation}
\begin{equation}
    \mathcal{B}_\infty=-\frac{1}{\pi}\sum_{\Delta_\varphi\leq\Delta_\mathcal{O}} \int_{|z|\leq\lambda^2}d^2z \frac{C_{\mathcal{O}_t W^{(1,1)}\varphi^c} C_{\mathcal{O}V^{(1,1)}\varphi}}{|z|^{\Delta_\varphi-\Delta_{\mathcal{O}}+2}}\,.
\end{equation}
Due to the fact that we are deforming the background by exactly marginal operators, $\cM_0$ can only contain polynomial divergences in $\lambda.$
On the other hand, $\cM_1$ and $\cM_\infty$ contain both the logarithmic and polynomial divergences. The logarithmic divergences are cancelled against \eqref{eqn:(1,1) log}. We will show that $\cB_0$ is cancelled by the vertical integration around $|z|\simeq\lambda^{-2}$, the polynomial divergence of $\cM_1$ is cancelled by the combined contribution of the vertical integration around $|z-1|\simeq\lambda^{-2}$ and the contribution from the Feynman region \cite{deLacroix:2017lif,Sen:2024nfd}
\begin{equation}\label{off-shell prop1}
   - \left\{ \mathcal{V}_t\otimes \Psi_{1,0}^0\otimes \frac{b_0^+}{L_0^+}(1-\Bbb{P})[\Psi_{0,1}^0\otimes \delta\Psi_{0,0}]\right\}\,,
\end{equation}
and finally the polynomial divergence of $\cM_\infty$ is cancelled by the combined contribution of the vertical integration around $|z|\simeq\lambda^2$ and
\begin{equation}\label{off-shell prop2}
  -  \left\{\mathcal{V}_t\otimes \Psi_{0,1}^0\otimes\frac{b_0^+}{L_0^+}(1-\Bbb{P})[\Psi_{1,0}^0\otimes\delta\Psi_{0,0}]\right\}\,.
\end{equation}

Let us first study the logarithmic divergences in $\cM_1$ and $\cM_\infty.$  Such divergences arise when $\varphi$ is $\mathcal{O}_t.$ Then, after an appropriate change of the integration variable we can combine  
\begin{equation}
    \mathcal{B}_{1,log}+\mathcal{B}_{\infty,log}=-2C_{S^2} \int_{|z|\leq\lambda^{2}}d^2z C_{OW^{(1,1)}\mathcal{O}_t}C_{\mathcal{O}_tV^{(1,1)}\mathcal{O}} |z|^{-2}=-4\pi C_{S^2}  C_{OW^{(1,1)}\mathcal{O}_t}C_{\mathcal{O}_tV^{(1,1)}\mathcal{O}}\log \lambda^{-2}\,,
\end{equation}
which perfectly cancels against \eqref{eqn:(1,1) log} as promised.

Now let us calculate \eqref{off-shell prop1} and \eqref{off-shell prop2}. We can write \eqref{off-shell prop1} as
\begin{equation}
   \frac{g_c^2}{4} \sum_\Phi \{ \mathcal{V}_t \otimes \Psi_{1,0}^0\otimes \Phi\} \left\langle \Phi^c\biggr|\frac{1}{L_0^+}(1-\Bbb{P})\biggr|\Phi\right\rangle\{ \Phi^c\otimes \Psi_{0,1}^0\otimes\delta\Psi_{0,0}\}\,.
\end{equation}
As we are taking the large stub limit, only the states 
\begin{equation}
    \Phi=c\bar{c}e^{-\phi}e^{-\bar{\phi}}\varphi e^{-ik\cdot x}\,,
\end{equation}
with $\Delta_\varphi <\Delta_\mathcal{O}$ contributes. Therefore, we conclude that \eqref{off-shell prop1} is equal to
\begin{equation}
    \label{eqn:off-shell1}2\pi\sum_{\varphi}C_{S^2} C_{\mathcal{O}_t V\varphi}C_{\mathcal{OW\varphi}} \frac{\lambda^{2\Delta_\mathcal{O}-2\Delta_\varphi}}{\Delta_\mathcal{O}-\Delta_\varphi}\,.
\end{equation}
Similarly, \eqref{off-shell prop2} is 
\begin{equation}
    \label{eqn:off-shell2}2\pi\sum_{\varphi}C_{S^2} C_{\mathcal{O}_t W\varphi}C_{\mathcal{OV\varphi}} \frac{\lambda^{2\Delta_\mathcal{O}-2\Delta_\varphi}}{\Delta_\mathcal{O}-\Delta_\varphi}\,,
\end{equation}
where we defined
\begin{equation}
    C_{\mathcal{O} W\varphi}:= \langle c\bar{c} e^{-\phi} e^{-\bar{\phi}} \mathcal{O}(0)\otimes c\bar{c} e^{-\phi}e^{-\bar{\phi}} W(1)\otimes c\bar{c} e^{-\phi}e^{-\bar{\phi}}\varphi(1)\otimes \mathcal{X}(p_1)\otimes \overline{\mathcal{X}}(p_2) \rangle\,,
\end{equation}
where $p_1$ and $p_2$ are the symmetric location of the PCOs. 

The polynomial divergences of $\mathcal{B}_1$ and $\mathcal{B}_\infty$ are not exactly canceling \eqref{eqn:off-shell1} and \eqref{eqn:off-shell2} because of the difference in the PCO locations used in the evaluation of the CFT correlation functions. However, the difference due to the PCO location is precisely what is provided by the vertical integration on the boundary of the moduli space. After taking into account the vertical integration, we conclude that the polynomial divergences from $\mathcal{B}_1$ and $\mathcal{B}_\infty$ are completely canceled.

Now we shall show that the vertical integration for $|z|=\lambda^{-2}$ cancels the polynomial divergence of $\mathcal{B}_0.$ This is of course expected. The effect of the vertical integration is to move one holomorphic and one anti-holomorphic PCO from the sphere with $W^{(1,1)},$ $V^{(1,1)}$ and $\varphi^c$ insertion to the other sphere. After the move of the PCOs, $\varphi^c$ should have a definite $U(1)_R$ charge both in the holomorphic and anti-holomorphic sector. As the collision of $W,~V$ and one holomorphic and one anti-holomorphic PCO can at best produce the singularity 
\begin{equation}
    |z|^{\Delta_\varphi-3}
\end{equation}
and due to the BPS condition that $\Delta_\varphi\geq1,$ after the move of the PCO only the logarithmic divergences remain. Therefore, one can conclude that the vertical integration completely cancels the polynomial divergence.

Let us briefly summarize the outcome and compare it to the naive perturbative CFT regularization scheme introduced in section~\ref{ss:naiveCPertT}. In the large $\lambda$ limit, the region $\cM$ in (\ref{eqn:4pt}) becomes $\C_\epsilon$ introduced in (\ref{Cepsilon}) with $\epsilon=\lambda^{-2}$. The terms coming from $\cA_2$ and $\cA_3$ and the vertical integration then play the role of the counterterms introduced in (\ref{naiveCT}). The upshot is then that the second order shift is determined by the constant part in the $\lambda$ expansion of the integral (\ref{eqn:4pt}), which is exactly the prescription given in section~\ref{ss:naiveCPertT}.

Even though the prescriptions agree, the advantage of the string field theory regularization scheme discussed here is that it is a scheme that is manifestly consistent to arbitrary order. The scheme introduced in section~\ref{ss:naiveCPertT} on the other hand is somewhat ad hoc. While it certainly works at second order, generalizing it to higher orders, though possible, introduces additional subtleties that need to be dealt with carefully so as to not introduce inconsistencies.

\section{Gepner models}\label{s:Gepner}
\subsection{$N=2$ minimal models: primaries and characters}\label{ss:bossubA}
Let us now introduce the ingredients of the CFT that we will consider. This section is based on \cite{Keller:2023ssv}, with slight modifications to adapt it to the case of the quintic. We mostly include it to make the article self-contained.
$N=2$ minimal models are CFTs that are rational with respect to the $N=2$ superconformal algebra. They have central charge
\be
c = \frac{3k}{k+2}
\ee
and can either be constructed a coset of WZW models or as a tensor product of $\Z_k$ parafermions with a free boson \cite{Fateev:1985mm,Zamolodchikov:1986gh}. We will focus on the case $k=3$. In the following we mostly follow the conventions of \cite{Stanishkov:2017rrr}, which provides a nice overview of $N=2$ minimal models.

In the NS sector, the primary fields are given by $N^l_m$, $l=0,1,\ldots k$ and $m=-l,-l+2,\ldots ,l$, with conformal weight and charge
\be
\Delta^l_m=\frac{l(l+2)-m^2}{4(k+2)}\ , \qquad Q^l_m = \frac{m}{k+2}\ .
\ee
In the R sector, the primary fields are given by $R^l_{m,\alpha}$, $l=0,1,\ldots k$, $m=-l,-l+2,\ldots ,l$ and $\alpha=\pm1$, with conformal weight and charge \footnote{Note our convention for the $U(1)$ charge differs by a factor of $\frac12$ from \cite{Stanishkov:2017rrr}. 
}
\be
\Delta^l_m=\frac{l(l+2)-(m+\alpha)^2}{4(k+2)}+\frac18\ , \qquad Q^l_m = \frac{m+\alpha}{k+2} -\alpha \frac12\ .
\ee
In view of the Gepner construction below, it is useful to decompose the irreducible representations of the $N=2$ SCA into irreducible representations of its bosonic subalgebra, that is the algebra generated by its bosonic generators and even numbers of fermionic generators. The primaries above then split into two irreducible representations with even and odd fermion number respectively. This will be very useful once we introduce the GSO projection, for which we need to keep track of the fermion parity.
Moreover, this also allows to treat Ramond and NS representations on equal footing, since they both have trivial monodromy with respect to the bosonic subalgebra. 

For labelling purposes we therefore introduce an additional parameter $s=-1,0,1,2$ that keeps track of the fermion parity and the Ramond and NS sectors. Here $s=0,2$ labels the two parities in the NS sector, and $s=-1,1$ in the R sector. The irreps of the bosonic subalgebra are thus labeled by triples $(l,m,s)$ with identifications $m\sim m\pm (2k+4)$ and $s\sim s\pm4$. Moreover we have the field identification
\be\label{fieldid}
(l,m,s) \sim (k-l,m+k+2,s+2)\ .
\ee
It can therefore be useful to have the labels run as
\be\label{labelcond}
0\leq l \leq k, \qquad m=-k-1,\ldots,k+2 \qquad s=-1,0,1,2\qquad s+m+l=0 \mod 2
\ee
and then by hand compensate for overcounting.
We then label the fields by $\Phi^l_{m,s;\bar m,\bar s}$, with weights and charge given by 
\be\label{hlms}
h^l_{m,s}=\frac{l(l+2)-m^2}{4(k+2)}+\frac{s^2}8 \mod 1\ ,
\qquad Q^l_{m,s}=\frac{m}{k+2}-\frac s2 \mod 2\ .
\ee
Their characters are given by \cite{Gepner:1987qi}
\be
\chi_m^{l(s)}(\tau,z)= \sum_{j\mod k} c^l_{m+4j-s}(\tau) \Theta_{2m+(4j-s)(k+2),2k(k+2)}(\tau,2kz,0)\ .
\ee
Here the classical theta function associated with $SU(2)$ at level $m$ is
\be
\Theta_{n,m}(\tau,z,u)= e^{-2\pi i u} \sum_{j\in \Z+\frac n{2m}} e^{2\pi \tau m j^2+2\pi i jz}
\ee
and $c^l_{m}$ is the string function of $A^{(1)}_1$,
\be
c^l_m(\tau) = \eta(\tau)^{-3} \sum_{\substack{-|x|<y\leq|x|\\ (x,y) \textrm{ or } (1/2-x,1/2+y)\in (\frac{l+1}{2(k+2)}, \frac{m}{2k})+\Z^2}} sign(x) \, e^{2\pi i\tau((k+2)x^2-ky^2)}\ .
\ee
These characters are indeed invariant under $s\mapsto s +4$, $m \mapsto m + 2k+4$ and under the identification $l\mapsto k-l, m\mapsto m+k+2, s\mapsto s+2$.

\subsection{The Quintic Gepner model: $(3)^5$}

We can now use these $\cN=2$ minimal models to construct a Gepner model. Gepner models were introduced in \cite{Gepner:1987qi,Gepner:1989gr}.  Their basic ingredient is a tensor product of $r$ minimal models such that their total central charge satisfies
\be
\sum_{i=1}^r \frac{3k_i}{k_i+2} = 3D\ .
\ee
Here $D$ is the complex dimension of the Calabi-Yau $D$-fold whose sigma model they represent, in our case  $D=3$. More concretely, we will focus on the Gepner model with $r=5, k_i=3$. This corresponds to the quintic Fermat surface.

The idea is then to construct the quintic sigma model partition function out of the characters of this tensor product. The natural starting point is simply the diagonal invariant of the tensor product of the bosonic subalgebra of the factors,
\be
Z= \sum_{(l_i,m_i,s_i)} \prod_{i=1}^5 \chi_{m_i}^{l_i(s_i)} (\chi_{m_i}^{l_i(s_i)})^*\ .
\ee
This partition function is clearly modular invariant, as it is a product of modular invariant functions. 
It does not however describe the CY sigma model that we are interested in. 
On the one hand, there is no proper notion of Ramond and NS sectors yet: in the tensor product, the individual factors are independently allowed to be in NS and Ramond sectors.
On the other hand, it does not contain the holomorphic top form $\Omega^{D/2,0}$ that acts as the (one unit) spectral flow operator. To construct the partition function of a consistent CFT, we want to perform a so-called simple current extension \cite{Schellekens:1989am,Schellekens:1990xy,Fuchs:1996dd}. The advantage of this relatively technical way of constructing Gepner models is that it not only guarantees a modular invariant partition function, but also allows to compute (consistent) correlators of the resulting theory.

Simple current extensions work in the following way:
A simple current $\Jsf$ is a holomorphic field whose fusion with any primary field $\varphi$ yields just a single field $\Jsf\varphi$. Single currents form a group $\cG$. We want to extend the chiral algebra of our CFT by such a $\Jsf$. To do this, we first project out any fields $\varphi$ whose monodromy charge
\be
Q_\Jsf(\varphi) := h_{\Jsf}+h_{\varphi}- h_{\Jsf \varphi}
\ee
is not integral. This ensures that $\Jsf$ is local, that is that the OPE of $\Jsf$ with $\varphi$ does not lead to branch cuts. The primary fields $\varphi$ that survive the projection are then organized into orbits $[\varphi ]$ of $\cG$. The diagonal modular invariant of the extended theory is
\be\label{Zext}
Z_{ext} = \sum_{\substack{[\varphi] \\ Q_\Jsf(\varphi)\in\Z\ \forall \Jsf\in\cG}}
|\cS_\varphi| \left| \sum_{\Jsf\in \cG/\cS_\varphi} \chi_{\Jsf\varphi}(\tau)\right|^2\ ,
\ee
where the $\chi$ are the characters of the original theory.
Here $\cS_\varphi$ is the subgroup of $\cG$ that leaves $\varphi$ invariant under the fusion product; we will return to it momentarily. Note that (\ref{Zext}) is non-diagonal with respect to the original characters. Off-diagonal states are often called \emph{twisted} sectors of the extension, even though they are strictly speaking not the same as twisted sectors of orbifold theories.

Let us now apply this to our minimal models.
We first want to extend the model by simple currents written schematically as $G^i G^j$. Here the indices denote in which factor the current lives, such as
\be
G^1G^2=N^0_{0,2;0,0}\otimes N^0_{0,2;0,0}\otimes N^0_{0,0;0,0}\otimes N^0_{0,0;0,0}\otimes N^0_{0,0;0,0}\ .
\ee
These form a group $\Z_2^4$. From (\ref{hlms}) we see that their projection implements $s_i/2+s_j/2\in\Z$, meaning that all factors are either in the NS or Ramond sector.
Including their twisted  sectors leads to a partition function that is a sum over terms
\be
\prod_{i=1}^5 \chi^{l_i(s_i)}_{m_i}(\chi^{l_i(\bar s_i)}_{m_i})^*\ ,
\ee
where the $s_i$ and $\bar s_i$ are all either odd or even (for the R-R or NS-NS sector respectively) and satisfy the GSO-type condition
\be\label{GSO}
\sum_{i=1}^5 \left(\frac{s_i}2+\frac{\bar s_i}2 \right) \in 2\Z\ \textrm{(in NS-NS) or}\ 2\Z+1\ \textrm{(in R-R)}\ .
\ee
Next, we want to extend by the operator 
\be
\Jsf= \bigotimes^5 N^0_{2,2;0,0} =  \prod_i e^{\frac i{\sqrt2}\phi_i(z)}\ . 
\ee
Following the notation in \cite{Recknagel:1997sb, Brunner:2006tc}, we label the twisted sectors that arise from this extension by an integer $n=0,1,2,3,4$. The character of the twist $n$ sector is then given by
\be\label{twistchar}
\prod_{i=1}^5 \chi^{l_i(s_i)}_{m_i+2n}(\chi^{l_i(\bar s_i)}_{m_i})^*\ ,
\ee
where the $s_i,\bar s_i$ satisfy the same conditions as above \cite{Fuchs:2000gv,Brunner:2004zd}.
In the NS sector, the condition is that the $U(1)$ charge is integral. In the R sector, the $U(1)$ charges are in $\frac D 2 + \Z$. For the quintic this leads to the condition
\be\label{mchargeintegral}
\sum_{i=1}^5 \frac{m_i}5 \in \Z \ \textrm{(in NS-NS) or}\ \Z+\frac12\ \textrm{(in R-R)}\ .
\ee
In total, the partition function of the quintic sigma model at the Gepner point $(3)^5$ is given by summing (\ref{twistchar}) over $l_i,m_i,s_i,\bar s_i,n$ subject to the conditions (\ref{labelcond}), (\ref{GSO}), (\ref{mchargeintegral})
and the identification (\ref{fieldid}).

\subsection{Expressions for the characters}\label{ss:charexp}
For concreteness, let us give the first few terms of the partition function in the NS-NS sector of our quintic Gepner models. In practice going to higher order becomes computationally very expensive very quickly. 
For the $n=0$ sector we get
\begin{multline}
Z_{n=0}=1+ 20 q^{1/5} \bar{q}^{1/5}+
115 q^{2/5} \bar{q}^{2/5}+101 \sqrt{q} y \sqrt{\bar{q}} \bar{y}+\frac{101 \sqrt{q} \sqrt{\bar{q}}}{y \bar{y}}+5 \bar{q}+5q +\ldots
\end{multline}
For $n=1$ we get
\be
Z_{n=1}=\frac{\sqrt{q} y \sqrt{\bar{q}}}{\bar{y}}+\ldots \ .
\ee
For $n=2$ and $n=3$ we get no contributions to this order.
Finally for $n=4$ we get
\be
Z_{n=4}= \frac{\sqrt{q} \bar y \sqrt{\bar{q}}}{y}+\ldots
+\ldots \ .
\ee
Collecting the five sectors, we indeed recover the Hodge diamond of the Quintic with 101 (c,c) and (a,a) moduli, and 1 (a,c) and (c,a) modulus. We also see that the lightest non-BPS states all come from the untwisted sector with total weights $2/5,4/5$ and 1.

In what follows we will only focus the 20 states of weight $(1/5,1/5)$ and charge 0. They are given by the tensor product of a factor $(l,m,s,\bar s)=(1,1,0,0)$, a factor $(1,-1,0,0)$ and three vacuum factors. There are indeed $5\times 4$ permutations of this, leading to 20 such states.

\section{Correlation functions}\label{s:corr}

\subsection{Minimal model correlators}
Let us now discuss how to compute the correlators of the $N=2$ minimal models. We start with correlation functions of $SU(2)$ WZW models. 
Let $\Phi^{(l)}_{m,\bar m}$ be the primaries of the the $SU(2)$ WZW model at level $k$. We can then obtain the parafermionic fields $\phi^{(l)}_{m,\bar m}$ from the $SU(2)$ fields $\Phi^{(l)}_{m,\bar m}$ via (4.31) in \cite{Mussardo:1988av} by splitting off a free boson,
\be
\Phi^{(l)}_{m,\bar m}(z,\zb)=\phi^{(2l)}_{2m,\bar 2m}(z,\zb) : e^{im\varphi(z)/\sqrt k + i\mb\bar \varphi(\zb) /\sqrt k}:\ .
\ee
This gives the four-point function
\be
\langle \prod_{i=1}^4 \phi^{(l_i)}_{m_i,\bar m_i}(z_i,\zb_i)\rangle
= \prod_{i<j}^4 z_{ij}^{-m_im_j/2k}\zb_{ij}^{-\mb_i\mb_j/2k}
\langle \prod_{t=1}^4 \Phi^{(l_t/2)}_{m_t/2,\bar m_t/2}(z_t,\zb_t)\rangle
\ee
Next, we connect these parafermions to $N=2$ minimal models. The level $k$ $N=2$ minimal model can be obtained by adjoining a free scalar $\varphi$ to the parafermion theory. This plays the role of the $U(1)$ current. The NS primaries $N^l_{m,\mb}$ of the minimal models are given by
\be
N^l_{m;\mb}(z,\zb) = \phi^{(l)}_{m,\mb}(z,\zb)
: e^{im\varphi(z)/\sqrt{2k(k+2)}  + i\mb\bar \varphi(\zb) /\sqrt{2k(k+2)}}:\ .
\ee
The supercharges act as
\be
G^\pm(z_1)N^l_m(z_2) = \sqrt{\frac{2k}{k+2}}\frac1{z_{12}}\phi^l_{m\pm 2}(z_2) e^{i(m\pm (k+2))/\sqrt{2k(k+2)}\varphi(z_2)}\ .
\ee
Finally, for the purpose of Gepner models, we define $N^l_{m,s;\mb,\bar s}$ via
\be
N^l_{m,0;\mb,0} = N^l_{m;\mb}\ ,
\qquad N^l_{m,\pm 2;\mb,0} =G^\pm_{-1/2}N^l_{m;\mb}\ ,
\qquad N^l_{m,0;\mb,\pm 2} =\bar G^\pm_{-1/2}N^l_{m;\mb}\ .
\ee
Overall this allows us to compute the four-point functions as 
\begin{multline}
\langle \prod_{i=1}^4 N^{(l_i)}_{m_i,s_i;\bar m_i,\bar s_i}(z_i,\zb_i)\rangle
\\= \prod_{i<j}^4 z_{ij}^{-\frac{m_im_j}{2(k+2)}+\frac{s_is_j}4}\zb_{ij}^{-\frac{\mb_i\mb_j}{2(k+2)}+\frac{\bar s_i \bar s_j}4}
\langle \prod_{t=1}^4 \Phi^{(l_t/2)}_{(m_t+s_t)/2,(\bar m_t+\bar s_t)/2}(z_t,\zb_t)\rangle K_{s_t}K_{\bar s_t}\ ,
\end{multline}
where the normalisation factor is
\be
K_{\pm2}=\sqrt{\frac{2k}{k+2}} \qquad \textrm{and}\qquad K_s=1 \textrm{ else.}
\ee
The upshot is that this allows us to compute minimal model correlation functions from the correlation functions of the $SU(2)$ WZW model. We will turn to these next.

\subsection{$SU(2)$ correlation functions}
Following \cite{Zamolodchikov:1986bd,Christe:1986cy} and also \cite{Mussardo:1988av}, let us now explain how to obtain the four-point functions of the $SU(2)$ primaries $\Phi^{(j)}_{m,\bar m}(z,\zb)$. 
The primary fields $\Phi^{(j)}_{m,\bar m}(z,\zb)$ have conformal weight
\be
\Delta^{(j)} = \frac{j(j+1)}{k+2}\ .
\ee
It is convenient to collect all fields of a given spin $j$ into one field. To this end we introduce an isospin variable $x$ and define the generating function
\be\label{Phixfield}
\Phi^{(j)}(x,\xb;z,\zb) = \sum_{m,\bar m = -j}^j \left(\binom{2j}{m+j}\binom{2j}{\bar m +j} \right)^{-1/2} x^{j+m}\xb^{j+\bar m} \Phi^{(j)}_{m,\bar m}(z,\zb)\ .
\ee
The four-point function of these fields are then given by \cite{Zamolodchikov:1986bd}
\begin{multline}\label{CorrFromU}
\langle \prod_{i=1}^4 \Phi^{(j_i)}(x_i,\xb_i;z_i\zb_i)\rangle = (x_{14}\xb_{14})^{2j_1}(x_{24}\xb_{24})^{-j_1+j_2-j_3+j_4}(x_{34}\xb_{34})^{-j_1-j_2+j_3+j_4}(x_{32}\xb_{32})^{j_1+j_2+j_3-j_4}\\
\times |z_{14}|^{2\nu_1} |z_{24}|^{2\nu_2} |z_{34}|^{2\nu_3}|z_{32}|^{2\nu_4}
U_{j_1j_2j_3j_4}(x,\xb;z,\zb)
\end{multline}
where
\bea
\nu_1&=& -2\Delta_1\\
\nu_2&=&\Delta_1-\Delta_2+\Delta_3-\Delta_4\\
\nu_3&=&\Delta_1+\Delta_2-\Delta_3-\Delta_4\\
\nu_4&=&-\Delta_1-\Delta_2-\Delta_3+\Delta_4\ ,
\eea
and the cross-ratios are
\be
z= \frac{z_{12}z_{34}}{z_{14}z_{23}}\ , \qquad x = \frac{x_{12}x_{34}}{x_{14}x_{23}}\ .
\ee
From this we can read off the four-point functions of the $\Phi^{(j)}_{m,\mb}$ by expanding in $x_i,\xb_i$.

Obviously the central object here is $U(x,\xb;z,\zb)$.  We will devote the next section and appendix~\ref{app:corr} to its computation. For now we can give it a physical interpretation by defining
\be
G_{1234}(x_i,z) = \lim_{w\to\infty} w^{2\Delta_2} \langle \Phi_2(w,x_2)\Phi_1(1,x_1)\Phi_3(z,x_3)\Phi_4(0,x_4)\rangle\ .
\ee
(Note the somewhat non-standard positioning of the fields due to the different convention for the cross-ratio.) We then have
\begin{multline}\label{GxzfromUxz}
G_{1234}(x_i,z) = \\
|z|^{2\nu_3}(x_{14}\xb_{14})^{2j_1}(x_{24}\xb_{24})^{-j_1+j_2-j_3+j_4}(x_{34}\xb_{34})^{-j_1-j_2+j_3+j_4}(x_{32}\xb_{32})^{j_1+j_2+j_3-j_4}U_{1234}(x,z)\ .
\end{multline}
We can thus interpret $U(x,\xb;z,\zb)$ as the correlation function (or, more precisely, the generating function for the correlation functions) with fields inserted at positions $z_1=1,z_2=\infty,z_3=z,z_4=0$.

\subsection{Obtaining $U$ from the FZ differential equation}
Let us now discuss how to calculate $U$.
Without loss of generality, we take $j_1\leq j_2\leq j_3\leq j_4$.
Next, we define the expansion 
\be\label{generalU}
U(x,\bar x;z,\zb) = \sum_{p,\bar p =p_0}^{p_1}(x-z)^p(\xb-\zb)^{\bar p}U_{p,\bar p}(z,\zb)
\ee
where
\bea
p_0 &=& \max(0,j_1+j_2+j_3+j_4-k)\\
p_1 &=& \min(2j_1,j_1+j_2+j_3-j_4)
\eea
Next, we can use an isospin version of the Knizhnik-Zamolodchikov equation to obtain a differential equation for $U$ \cite{Zamolodchikov:1986bd}. This isospin version also contains the variable $x$ and its derivative. In terms of the expansion (\ref{generalU}), it just leads to a system of differential equations for the
functions $U_{p,\bar p}$:
\begin{multline}\label{Udiff}
(k+2)z(z-1)\frac{\partial}{\partial z}U_{p,\bar p} =
(p+1)(p+1+k-J)z(z-1)U_{p+1,\bar p} \\
- (a_pz+b_p(z-1))U_{p,\bar p} + (2j_1-p+1) (J+1-2j_4-p)U_{p-1,\bar p}
\end{multline}
where
\be\label{apbp}
J = j_1+j_2+j_3+j_4 \qquad a_p = p(p+1)-2(p-j_1)(p-j_3)\qquad b_p = p(p+1) - 2(p-j_1)(p-j_2)\ .
\ee
There are of course analogous equations for $\zb$. 

Let us denote by $v_p^{(j)}$ a basis of solutions for (\ref{Udiff}), where the vector index $p$ runs from $p_0$ to $p_1$. We will interpret $j$ as the spin of the primary running in the internal channel, so that
\be
\max(j_2-j_1,j_4-j_3) \leq j \leq \min(j_1+j_2,k-j_3-j_4)\ .
\ee
Indeed we find that we can write 
\be
v_p^{(j)}(z) = z^{\kappa_j}f_p^{(j)}(z) \ , 
\ee
where the $f_p^{(j)}(z)$ are regular functions at $z=0$. We will fix the normalization as  $f_{p_1}^{(j)}(0)= 1$.
Here the order of the singularity $\kappa_j$ is determined by the weights $\kappa_j = \Delta^{(j)}-\Delta^{(j_1)}-\Delta^{(j_2)}$. This is thus indeed compatible with the value of $j$ of the primary that runs in the internal channel. The $v^{(j)}_p(z)$ are thus the conformal blocks.

To obtain the correlator, we write the usual sesquilinear combination of conformal blocks
\be
U_{p,\bar p}(z,\zb) = \sum_j A_j v_p^{(j)}(z) v_{\bar p}^{(j)}(\zb)\ .
\ee
This ensures that the correlator is single-valued around $z=0$.
The coefficients are given by the structure constants
\be\label{Ajfrom3pt}
A_j = C(j_1,j_2,j)C(j_3,j_4,j)\ .
\ee
In total we have
\begin{multline}\label{Uexpression}
U(x,\xb;z,\zb) = \sum_{p,\bar p=p_0}^{p_1} (x-z)^p(\xb-\zb)^{\bar p} U_{p,\bar p}(z,\zb)\\
= \sum_j A_j \left(\sum_{p} (x-z)^p v_p^{(j)}(z)\right) \left(\sum_{\bar p} (\xb-\zb)^{\bar p} v_{\bar p}^{(j)}(\zb)\right)
\end{multline}
It turns out that for our specific computations, the system (\ref{Udiff}) is relatively simple: We only need the cases with one or two internal channels.
Note that
\be
p_1-p_0 \leq k -2j_4
\ee
and
\be
p_1-p_0 \leq 2j_1\ .
\ee
It follows that for $j_4=k/2$, $p_0=p_1$, so that $p$ and $\bar p$ only take 1 value. The system thus reduces to a single first order differential equation, which has an elementary solution. For $j_4=(k-1)/2$ or $j_1=1/2$, they only take 2 values. In those cases we can combine the two first order differential equation into a single second order differential equation. This differential equation turns out to be a hypergeometric differential equation, so that the solutions are hypergeometric functions. This allows us to write down explicit expressions for the $v_p^{(j)}$. We can then use crossing symmetry to fix the coefficients $A_j$.

The case $j_1=1/2$ was worked out explictly in \cite{Zamolodchikov:1986bd}. We give the expressions in appendix~\ref{ss:U12}. We will also need the case $j_4=k/2$, which we work out in appendix~\ref{ss:Uk2}.

\section{Lifting the lightest states at $h=(1/5,1/5)$}\label{s:shift}

\subsection{Moduli and states}
Now that we have set up all the ingredients, let us finally compute the corrections to K\"ahler metric. To compute the full one-loop correction to the K\"ahler metric as in (\ref{eqn:torus two2}), we would have to compute the shift of the weight of all the states in the spectrum of the theory. We will be less ambitious here and instead only compute the shift of the lightest non-protected states that we described at the end of section~\ref{ss:charexp}. We will also not evaluate the integral (\ref{eqn:torus two2}). Even though the lightest states whose lifting we compute will presumably make the most important contribution to the integral, heavier states may make significant contributions as well, so that the integral of just the lightest states would not give reliable information. We leave heavier states and evaluating the integral to a future project.

First, let us describe the moduli that we found in section~\ref{ss:charexp} in more detail.
The $k=3$ minimal model has chiral fields $N^l_l$ with $h=l/10$ and $q=l/5$, and anti-chiral fields $N^l_{-l}$ with $h=l/10$ and $q=-l/5$, with $l=0,1,2,3$. In the language of Fermat polynomials these correspond to monomials $x_i^l$. Moduli are then given by polynomials of total degree 5 with individual degrees less or equal to 3. For concreteness we will consider perturbation by the term $x_1^3x_2^2$. 
On the CFT side, this corresponds to the modulus given the by $G$ descendant 
\be\label{Omodulus}
O = G_{-1/2}\bar G_{-1/2}\left( N^3_{3;3}\otimes N^2_{2;2}\otimes 1\otimes 1\otimes 1\right)\ .
\ee
Here $G$ denotes the diagonal supercharge of all the tensor factor supercharges, that is
\be\label{Gdiag}
G\otimes 1 \otimes 1 \otimes 1\otimes 1 +1\otimes G \otimes 1 \otimes 1\otimes 1 + 1\otimes 1 \otimes G \otimes 1 \otimes 1 +1\otimes 1 \otimes 1 \otimes G \otimes 1+1\otimes 1 \otimes 1\otimes 1 \otimes G\ ,
\ee
and the $N=1$ supercharge itself is given by 
\be
G=\frac{G^++G^-}{\sqrt{2}}\ .
\ee
When acting on (anti-)chiral primaries however only one of the $N=2$ supercharges survives. Note that even though $G$ is not in the theory due to the GSO projection, $G\bar G$ is. 

It is useful to return to the bosonic subalgebra notation introduced in section~\ref{ss:bossubA}. The chiral primary is then
\be
N^l_{l,0;l,0}
\ee
and we have
\be
G^-_{-1/2}N^l_{l,0;l,0} \sim N^l_{l,-2;l,0}\ ,
\qquad G^+_{-1/2}N^l_{l,0;l,0} =0\ .
\ee

Finally let us discuss the states whose lifting we will compute. We will consider the states of weight $(1/5,1/5)$ and charge 0. As mentioned above, there are 20 such states, consisting of one factor with $N^1_{1;1}$, one factor with $N^1_{-1;-1}$, and three factors with the vacuum. We will denote them by 
\be
\varphi_{ij} \qquad i\neq j\  ,\  i,j=1,\ldots 5 \ ,
\ee
where $i$ is the position of the factor with $N^1_{1;1}$ and $j$ the position of the factor with $N^1_{-1;-1}$.

\subsection{The first order lifting matrix}
Since there are 20 states of the same weight, we need to perform degenerate perturbation theory to deal with possible operator mixing. Concretely this means we need to compute the $20\times20$ lifting matrix and then diagonalize it to obtain the shift of the weights.

Let us first discuss this at first order. Factorwise charge conservation makes most of the entries vanish. By standard first order perturbation theory, the entries of the lifting matrix under perturbation by a modulus $\Phi$ are then given by the three-point functions
\be
-\pi C_{\varphi^\dagger_{ij}\Phi\varphi_{kl}}\ .
\ee
First consider $O$, given by (\ref{Omodulus}).
We note that $N^3_{3,s}$ has charge $3/5+s/2$.  The only way to satisfy charge conservation in the first factor is to have $j=1,k=1$ and $s=-2$.
For the second factor, $N^2_{2,s}$ has charge $2/5+s/2$. The only way to satisfy charge conservation is to have $i=2,l=2$ and $s=0$. The only non-vanishing entry in the lifting matrix thus comes from
\be
C_{\varphi_{21}^\dagger O \varphi_{12}}= C_{\varphi_{12}O \varphi_{12}}\ .
\ee
We note that the lifting matrix is not hermitian and can therefore not be diagonalized to determine the lifting. This is not surprising, since $O$ is not a hermitian operator. We therefore also consider $O^\dagger$. For this the same argument as above holds, with the only non-vanishing entry coming from
\be
C_{\varphi_{12}^\dagger O^\dagger \varphi_{21}}= C_{\varphi_{21}O^\dagger \varphi_{21}} = C_{\varphi_{12}O \varphi_{12}}\ .
\ee
The full lifting matrix is thus reduced to a $2\times 2$ matrix.
We can then consider the two hermitian moduli
\be
\Phi_1 = \frac{O+O^\dagger}{\sqrt 2}\ , \qquad \Phi_2 = i\frac{O-O^\dagger}{\sqrt 2}\ 
\ee
with real coupling constants $\mu_1$ and $\mu_2$. 
Their lifting matrices are then 
\be
-\frac{\pi}{\sqrt 2} \begin{pmatrix} 0 & C_{\varphi_{12}O \varphi_{12}}\\
C_{\varphi_{12}O \varphi_{12}} & 0 \end{pmatrix}\ , \qquad
-\frac{\pi}{\sqrt 2} \begin{pmatrix} 0 & iC_{\varphi_{12}O \varphi_{12}}\\
-iC_{\varphi_{12}O \varphi_{12}} & 0 \end{pmatrix}.
\ee
All other matrix entries vanish. We thus have states with
\be
h^\pm(\mu_1,\mu_2) = \frac1{10} \pm \frac{\pi C_{\varphi_{12}O \varphi_{12}}}{\sqrt 2} \sqrt{\mu_1^2+\mu_2^2}+\ldots\ .
\ee
The corresponding parameters for $O$ and $O^\dagger$ are given by 
\be
\mu_1 = \frac{\mu+\bar \mu}{\sqrt 2}\ , \qquad \mu_2 = \frac{-i\mu+i\bar \mu}{\sqrt 2}\ .
\ee
Returning to (\ref{Kaehlercontribution}), we can use this to compute the derivatives $\partial h^\pm$ and $\bar \partial h^\pm$. In fact, because
\be
h^\pm(\mu, \bar\mu=0) = \frac{1}{10}+0+\ldots\ , \qquad h^\pm(\mu=0, \bar\mu) = \frac{1}{10}+0+\ldots \ ,
\ee
it follows that 
\be
\partial h = \bar \partial h= 0\ .
\ee
We see that the first order contributions vanish. Let us therefore turn to the second order contributions.

\subsection{The second order lifting matrix}
Let us now compute the matrix elements $\langle ij|O O^\dagger|kl\rangle$ of the lifting matrix of the states $\varphi_{ij}$ at second order.
As discussed in section~\ref{ss:naiveCPertT}, these elements
come from the constant term of the regularized integral of the 4pt function
\be\label{4ptcorr}
\langle \varphi_{ij}^\dagger(\infty) O(1)O^\dagger(x)\varphi_{kl}(0)\rangle\ .
\ee
First we note that a priori, each entry has $2^2\times2^2=16$ terms coming from the form (\ref{Omodulus}) and (\ref{Gdiag}) of the moduli. To keep track of those terms, it is useful to use the bosonic notation $N^3_{3,s;3,\bar s}$ and $N^2_{2,s;2,\bar s}$, where $s$ and $\bar s$ keep track of the $G$ and $\bar G$ descendants. 
We then observe that in each factor, the total $U(1)$ charge of the two moduli is integer: In the first factor, the moduli are $N^3_{3,s}(1)N^3_{-3,s'}(x)$ giving $U(1)$ charge $(s+s')/2$ (where $s,s'$ are 0 or $\pm 2$). The same also holds for the second factor with $N^2_{2,s}(1)N^2_{-2,s'}(x)$, and the remaining three factors trivially have zero $U(1)$ charges from the moduli. Since $\varphi_{kl}$ has $U(1)$ charge $1/5$ and $-1/5$ in the factors $k$ and $l$, the only way to satisfy charge conservation in every factor is to pair it with the state $\varphi_{kl}^\dagger$. 

The first immediate consequence of this is that the lifting matrix is diagonal. The second consequence is that the moduli actually have to give vanishing charge in each sector, meaning that $s=-s'$ and similarly for the right movers in each factor. This implies that of the 16 $G_{-1/2}$ descendant terms in (\ref{Gdiag}), only four can be non-vanishing, namely
\bea
\langle ij|N^3_{3,0;3,0}\otimes N^2_{2,-2;2,-2}(1)N^3_{-3,0;-3,0}\otimes N^2_{-2,2;-2,2}(x)|kl\rangle &&\\
\langle ij|N^3_{3,-2;3,0}\otimes N^2_{2,0;2,-2}(1)N^3_{-3,2;-3,0}\otimes N^2_{-2,0;-2,2}(x)|kl\rangle &&\\
\langle ij|N^3_{3,0;3,-2}\otimes N^2_{2,-2;2,0}(1)N^3_{-3,0;-3,2}\otimes N^2_{-2,2;-2,0}(x)|kl\rangle &&\\
\langle ij|N^3_{3,-2;3,-2}\otimes N^2_{2,0;2,0}(1)N^3_{-3,2;-3,2}\otimes N^2_{-2,0;-2,0}(x)|kl\rangle &&
\eea
Next, let us discuss how to compute the 16 diagonal entries of the lifting matrix. The correlation function factorizes into five factors. The three last ones are at most two-point functions, do not depend on $x$ and are therefore normalized to be 1. The first two factors are then either two-point functions of just the two moduli, or four-point functions of two moduli with $l=3$ or $l=2$ and two primaries with $l=1$.
For the four-point functions, in terms of the $SU(2)$ WZW correlation functions described in section~\ref{s:corr}, the first factor has $j_1=j_2=1/2$ and $j_3=j_4=3/2$, so that we are in the $j_4=k/2$ case. The second factor has $j_1=j_2=1/2$ and $j_3=j_4=1$, so that we are in the $j_1=1/2$ case. To compute the correlation function (\ref{4ptcorr}), it is thus enough to know the correlation functions that are worked out in appendices \ref{ss:Uk2} and \ref{ss:U12}.

\subsection{Computing the second order integral}
Now that we have established how to obtain the four-point function, let us next discuss how to evaluate the integral
\be
\int_{\C_\epsilon} d^2x \langle \varphi_{ij}^\dagger(\infty) O(1)O^\dagger(x)\varphi_{ij}(0)\rangle\ .
\ee
We note that the integrand has a non-integrable divergence at $x=1$ when $O^\dagger$ collides with $O(1)$ to produce the vacuum. For computational reasons it is more convenient to have the highest singularities at 0 or $\infty$. We therefore perform a change of variable $x\mapsto 1-x$, turning the integrand into
\be
I(x) = \langle \varphi_{ij}^\dagger(\infty) \varphi_{ij}(1)O^\dagger(x)O(0)\rangle\ .
\ee
We now want to regularize the integral of $I(x)$ and evaluate it. This is worked out in detail in \cite{Keller:2023ssv}, so we will only discuss the most relevant aspects here. 

As discussed in section~\ref{ss:naiveCPertT}, we use a hard sphere regularization scheme, that is we cut out discs of radius $\epsilon$ around the insertion points $0,1,\infty$ and integrate over the remaining region $\C_\epsilon$. We then evaluate the integral numerically by going to polar coordinates. Polar coordinates of course make it easy to implement the $\epsilon$-discs around $0$ and $\infty$, but hard for the $\epsilon$-disc around $1$. We therefore use a trick to evaluate the integral: We first subtract from $I(x)$ its non-integrable singularities at $x=1$,
\be
I^r(x) = I(x) - I^{sing}(x)\ ,
\ee
where 
\be\label{Ising}
I^{sing}(x)= \sum_{h+\hb \leq 2}\frac{A_{h,\hb}}{(x-1)^h(\xb-1)^\hb}\ .
\ee
We then write the integral of $I^r(x)$ in polar coordinates, 
\be\label{Irpolar}
\int_{\C_\epsilon}d^2x I^r(x) = \int_\epsilon^{\epsilon^{-1}} dr \int_0^{2\pi} I^r(r,\theta) d\theta +o(1)\ .
\ee
The $o(1)$ term appears because on the RHS of (\ref{Irpolar}), we neglected to cut out the $\epsilon$-disc around 1. However, since $I^r(x)$ by construction is integrable at $x=1$, the difference is $o(1)$, \ie goes to 0 as $\epsilon \to 0$. Similarly we evaluate the contribution of $I^{sing}(x)$ as
\be\label{Isingpolar}
\int_{\C_\epsilon}d^2x I^{sing}(x) = \int_\epsilon^{\epsilon^{-1}} dr \int_0^{2\pi} I^{sing}(1-x) d\theta +o(1)\ .
\ee
Here we again get a $o(1)$ correction term from neglecting the $\epsilon$-disc at $x=0$ and using the fact that $I^{sing}(x)$ is regular at $x=0$. In total, the regularized integral $\int_{\C_\epsilon} d^2 x I(x)$ up to $o(1)$ terms is thus given by the sum of (\ref{Irpolar}) and (\ref{Isingpolar}). 

Since $I^{sing}(x)$ is of the simple form (\ref{Ising}), evaluating $(\ref{Isingpolar})$ is straightforward. To evaluate (\ref{Irpolar}), we first split it up as
\be
\int_\epsilon^1 dr \int_0^{2\pi} I^r(x) d\theta +  \int_1^{\epsilon^{-1}} dr \int_0^{2\pi} I^r(x) d\theta\ .
\ee
To perform the first integral, we expand $I(x)$ in $r$ around 0, keeping all terms up to a given order. It is then straightforward to perform the $\theta$ and $r$ integrals order to order in $r$. Although in principle we could keep analytic expressions at each order, it is much faster to do the expansion numerically. Going to high enough order in $r$, the results then converge.
For the second integral we follow the same basic procedure, but we need to be a bit more careful because the underlying hypergeometric functions have branch cuts from 1 to $\infty$, so that the single-valuedness of the correlator is not always manifest. To get the correct expansion, we first use the identity
\be
F\left(a,b;c;\frac z{z-1}\right) = (1-z)^aF(a,c-b;c;z) \qquad arg(1-z) \neq \pi \ .
\ee
This identity holds if $1-z$ stays away from the negative real line. To expand around infinity, we therefore want to expand in $-z$ rather than $z$. We then convert to polar coordinates and perform the integral term by term up a given order as before.

\subsection{Results}
We performed this procedure for an expansion up to order $O(r^{12})$ using Mathematica. The corresponding notebook can be found as an ancillary file. We obtained the following entries for the diagonal matrix elements:
\begin{description}
\item[$i=1,j=2$:]
\be\label{i1j2}
\frac{2 \pi}{\epsilon ^2} - \gamma_1\frac{1}{\epsilon^{6/5}}+\gamma_2 \log (\epsilon )-0.90 + \ldots
\ee
\item[$i=2, j=1$:]
\be
\frac{2\pi}{\epsilon ^2}- \gamma_1\frac{1}{\epsilon^{6/5}}+\gamma_2 \log (\epsilon )+0.56 + \ldots
\ee

\item[$i=1,j>2$:]
\be
-\frac{ \pi}{9}\log (\epsilon )-0.87 +\ldots
\ee

\item[$i>2,j=1$:]
\be
-\frac{ \pi}{9} \log (\epsilon )-0.13 +\ldots
\ee
\item[$i=2,j>2$:]
\be
- \gamma_1\frac{1}{\epsilon^{6/5}}-\frac{ \pi}{9} \log (\epsilon )+0.60 +\ldots
\ee

\item[$i>2,j=2$:]
\be\label{ig2jg2}
- \gamma_1\frac{1}{\epsilon^{6/5}}-\frac{ \pi}{9} \log (\epsilon )+0.53 +\ldots
\ee

\item[$i,j>2$:]
In this case the correlation function factorizes into a product of two-point functions. Our regularization scheme therefore does not give a constant term, so that the entry is 0.

\end{description}
In the above, the constants $\gamma_i$ are
\be
\gamma_1 = \frac{\sqrt{5}\, 2^{1/5} \pi^2}{12} \frac{\Gamma(3/10)^2}{\Gamma(2/5)^4 }\ ,
\qquad \gamma_2 = \gamma_1 - \frac{4\pi}{225}\ .
\ee
The terms that are singular in $\epsilon$ are then cancelled by the counterterms. We note that the terms that diverge logarithmically actually come from the quadratic contribution of first order perturbation theory \cite{Keller:2023ssv}. We found that the constant terms have converged fairly well at order $O(r^{12})$, so that we did not try to push to higher order.

Following the logic described in \eg \cite{Keller:2023ssv}, we find that
\be
\partial \bar\partial h_i = -\pi M_{ii}\ ,
\ee
where $M_{ii}$ is the constant part of the expressions above.

Let us summarize our results. The constant terms in (\ref{i1j2}) through (\ref{ig2jg2}) give the diagonal entries of the second order lifting matrix for the 20 lightest states of weight $h=(1/5,1/5)$. Through (\ref{Kaehlercontribution}) this gives the contribution of these states to the change in the partition function. To compute the corrections to the K\"ahler potential through the derivatives of the string-threshold corrections (\ref{Deltaa}), one would first have to repeat this analysis for all other states contributing to the partition function. The main contribution will come from the light states, or more precisely in view of (\ref{Zh}) and (\ref{Zhbar}), the states with lowest Ramond sector weight. Unfortunately already the number of next to lightest states is in the thousands, so that we did not attempt to compute their contribution. To obtain the threshold corrections $\Delta_a$, one would then also have to evaluate the modular integral in (\ref{Deltaa}).

\section*{Acknowledgments}
The work of CAK was supported in part by NSF Grant 2111748. CAK thanks ETH Z\"urich and the Albert Einstein Center for Fundamental Physics at the University of Bern for hospitality, where part of this work was completed.

\appendix

\section{String vertices}\label{app:stringvertices}

\subsection{Sphere with three punctures}
In this appendix we determine vertex regions for sphere three and four-point string vertices and fix the corresponding local coordinates.
Sphere diagrams with three punctures have zero moduli. We shall, therefore, fix the locations of the punctures at $0,~1,$ and $\infty.$ We choose the local coordinates, as in \cite{Sen:2019jpm}, 
\begin{equation}
    w_i=\lambda f_i(z)\,,
\end{equation}
where
\begin{equation}
    f_1(z)= \frac{2z}{z-2}\,,\quad f_2(z)=-2\frac{1-z}{1+z}\,,\quad f_3(z)=\frac{2}{1-2z}\,.
\end{equation}
We choose to insert PCOs at the symmetric location
\begin{equation}
    p_\pm =\frac{1}{2}\pm i\frac{\sqrt{3}}{2}\,.
\end{equation}
However, when all three punctures are inserted with on-shell vertices, we will move PCOs to locations that are convenient for calculations.

\subsection{Sphere with four punctures}\label{app:4ptsphere}
Sphere diagrams with four punctures have a two dimensional moduli space which decomposes into a vertex region and three Feynman regions. As all four-point functions we shall compute in this paper only concern on-shell primary vertex operators, we do not need the details of the local coordinates in the vertex region. However, to properly regulate the divergences in the degeneration limit, we must know the details of the Feynman regions. Therefore, we shall determine Feynman regions using the plumbing fixture to join two three punctured spheres.

Let us denote the global coordinate of the four-punctured sphere by $z$, and the coordinates of the three punctured spheres by $x_1$ and $x_2.$ Without loss of generality, we shall glue the punctures at $x_1=x_2=0$ via
\begin{equation}
    \lambda^2 f_1(x_1)f_1(x_2)=q\,,
\end{equation}
where $q:= e^{-s+i\theta}$ is the modulus in Schwinger parametrization. 

The plumbing fixture maps $x_2$ to $x_1$ via
\begin{equation}
    x_1=\frac{2q\lambda^{-2}f_1(x_2)^{-1}}{2+q\lambda^{-2}f_1(x_2)^{-1}}\,.
\end{equation}
The punctures at $x_2=0,~1,$ and $\infty$ are mapped to
\begin{equation}
    x_1=2,\quad \frac{-2q\lambda^{-2}}{4-q\lambda^{-2}}\,,\quad \frac{2q\lambda^{-2}}{4+q\lambda^{-2}}\,,
\end{equation}
respectively. By relating $z$ to $x_1$ via an $SL(2;C)$ transformation, we can bring points at $x_1=1,~\infty,$ and $x_2=1,~\infty$ to
\begin{equation}
    z=1,\quad \infty,\quad 0,\quad \frac{16q\lambda^{-2}}{(4+q\lambda^{-2})^{2}}\,.
\end{equation}
Therefore, the boundary of the Feynman region is where the location of the movable puncture is restricted to
\begin{equation}
    |z|= \frac{16\lambda^{-2}}{|(4+e^{i\theta}\lambda^{-2})|^2}\,.
\end{equation}

\section{Correlation functions}\label{app:corr}

\subsection{Conformal blocks for $j_1=1/2$}\label{ss:U12}
In this section, we describe how to obtain the conformal blocks for the case $j_1=1/2$. This material can be found in appendix A of \cite{Zamolodchikov:1986bd}, where the result was obtained by evaluating a multiple contour integral. We include it for completeness and also to show how the results can be obtained from bootstrapping the 4pt function instead; 
this we will discuss in the next section.

For $j_1=1/2$ we have $p_0=0, p_1=1$.
We have the system of equations
\bea
(k+2)z(z-1)\frac{\partial}{\partial z}U_{0}(z) &=&
(j_2(z-1)+j_3 z) U_0(z) + (1-j+k)(z-1)zU_1(z) \label{eqj1half1} \\
(k+2)z(z-1)\frac{\partial}{\partial z}U_{1}(z) &=&
-((1+j_3)z+(1+j_2)(z-1))U_{1}(z) +(j-2j_4)U_0(z)
 \label{eqj1half2}
\eea
As described above, we can combine the system into a single second-order differential equation, which has hypergeometric functions as solutions. 
More precisely, we solve (\ref{eqj1half2}) for $U_0(z)$ and plug this into (\ref{eqj1half1}) to get a second order differential equation for $U_1(z)$. We then make the ansatz 
\be
U_1(z) = z^r(1-z)^s f(z)\ ,
\ee
and choose $r,s$ such that the differential equation turns into the hypergeometric differential equation,
\be
z(1-z) f''(z) + (c -(a+b+1)z)f'(z) -ac f(z)= 0\ ,
\ee
which has the solution ${}_2F_1(a,b;c;z)$. $U_0(z)$ can then be obtained by plugging the solution for $U_1(z)$ into (\ref{eqj1half2}).

The $r,s$ are determined by a quadratic equation, giving two independent solutions corresponding to the two internal channels with $j= j_2\pm 1/2$. The first solution is 
\bea
v_1^+(z) &=& (1-z)^{\frac{j_3}{k+2}}z^{\frac{j_2}{k+2}}F(a,b;c;z)
\eea
\begin{multline}
    v_0^+(z) = \frac{ (1-z)^{\frac{j_3}{k+2}} z^{\frac{j_2}{k+2}} } {\hf +j_2+j_3-j_4} \Big( \left( 2 j_2 (z-1) + 2 (j_3 +1)z -1  \right)F(a,b;c;z) \\
    + ab (k+2) z(z-1) F(a+1,b+1;c+1;z)\Big)
\end{multline}
where
\bea
a &=& \frac{j_2+ j_3+ j_4+3/2}{ k+2} \\
b &=& 1 +\frac{ j_2+ j_3- j_4+1/2}{ k+2}\\ 
c &=& 1+ \frac{2 j_2+1}{k+2}.
\eea
The second solution has $r= -\frac{j_2+1}{k+2}, s=\frac{j_3}{k+2}$, giving
\bea
v_1^-(z) &=& (1-z)^{\frac{j_3}{k+2}}z^{-\frac{j_2+1}{k+2}}F(a',b';c';z)\ ,
\eea
\begin{multline}
    v_0^-(z) = \Gamma \left(c'\right) \frac{z^{1-\frac{j_2+1}{k+2}} (1-z)^{\frac{j_3}{k+2}} }{ ( j_2+ j_3- j_4+1/2)} \Big( (2 j_3+1) \, _2F_1\left(a',b';c';z\right)
\\
+a'b'(k+2) (z-1)_2F_1\left(a'+1,b'+1;c'+1;z\right)\Big)
\end{multline}
where now 
\bea
a'&=& \frac{ -  j_2 +  j_3 +  j_4 + 1/2}{k+2}\\
b'&=& 1- \frac{ j_2 -  j_3 +  j_4 +1/2}{k+2}\\
c'&=& 1- \frac{2j_2 +1}{k+2}.
\eea
We can compare this to the expressions in appendix A of \cite{Zamolodchikov:1986bd}.
A straightforward computation shows that 
\bea\label{vFone}
v^\pm_1(z) &=& z^{\frac{j_2}{k+2}} (1-z)^{\frac{j_3}{k+2}}\cF^\pm_1(z)\\
\label{vFzero}
-z v_1^\pm(z) + v_0^\pm(z)&=&z^{\frac{j_2}{k+2}} (1-z)^{\frac{j_3}{k+2}}\cF^\pm_0(z)\ ,
\eea
where the $\cF^{(\sigma)}_i$ are given in (A2) in \cite{Zamolodchikov:1986bd}.
Plugging this into (\ref{Uexpression}) we indeed obtain
\begin{multline}\label{UexpA}
U(x,\xb;z,\zb) = \sum_{\sigma=\pm} A_\sigma \left(v_0^{(\sigma)}(z)-z v_1^{(\sigma)(z)}+x v_1^{(\sigma)}(z)\right)
\left(v_0^{(\sigma)}(\zb)-\zb v_1^{(\sigma)(\zb)}+\xb v_1^{(\sigma)}(\zb)\right)\\
= |z|^{\frac{2j_2}{k+2}}|1-z|^{\frac{2j_3}{k+2}}\sum_{\sigma=\pm} A_\sigma \left(\cF_0^{(\sigma)}(z)+x \cF_1^{(\sigma)}(z)\right)
\left(\cF_0^{(\sigma)}(\zb)+\xb \cF_1^{(\sigma)}(\zb)\right)
\end{multline}
where we used (\ref{vFone}) and (\ref{vFzero}). This reproduces (A1) in \cite{Zamolodchikov:1986bd}.

\subsection{Bootstrapping $U_{\frac12\frac12\frac12\frac12}$}
The coefficients $A_\pm$ in the four-point function (\ref{UexpA}) are obtained  in \cite{Zamolodchikov:1986bd} by relating the four-point function to an integral. Here we will explain how to obtain them by bootstrapping and crossing symmetry.

If we want to check crossing symmetry under $z\mapsto 1-z$, we have
\be\label{Gcross}
G_{1234}(x_i,z) = G_{4231}(x_{\sigma(i)},1-z)\ ,
\ee
where $\sigma$ is the transposition $(14)(2)(3)$.
Translating (\ref{Gcross}) into the crossing symmetry condition for $U(x,z)$, we obtain 
\be\label{Ucross}
U_{4231}(1-x,1-z) = \frac{|z|^{2(\Delta_1+\Delta_2-\Delta_3-\Delta_4)}}{|1-z|^{2(\Delta_4+\Delta_2-\Delta_3-\Delta_1)}} (x\xb)^{-j_1-j_2+j_3+j_4}((1-x)(1-\xb))^{-j_1+j_2-j_3+j_4} U_{1234}(x,z)
\ee
Note that in the case of $j_1=j_2=j_3=j_4$, this simplifies to 
\be\label{Ucross12}
U_{4231}(1-x,1-z)=U_{1234}(x,z)\ .
\ee
Specializing the expressions for the $v(z)$ in section~\ref{ss:U12} to all $j_i=1/2$, we find the crossing transformations
\bea\label{v1mcross}
v^-_1(z) &=&  \beta v^-_1(1-z) +\alpha v^+_1(1-z) \\
v^+_1(z) &=& -\beta  v^+_1(1-z) + \gamma v^-_1(1-z)\\
v^-_0(z) &=& -\beta  v^-_0(1-z) - \alpha v^+_0(1-z)\\
v^+_0(z) &=& \beta v^+_0(1-z) -\gamma v^-_0(1-z) \label{v0pcross}
\eea
where
\bea
\alpha &=& \frac{2 \Gamma \left(-\frac{2}{k+2}\right)^2}{3 \Gamma \left(-\frac{3}{k+2}\right) \Gamma \left(-\frac{1}{k+2}\right)}\\
\beta  &=& \frac{2 \Gamma \left(-\frac{2}{k+2}\right) \Gamma \left(\frac{2}{k+2}\right)}{\Gamma \left(-\frac{1}{k+2}\right) \Gamma \left(\frac{1}{k+2}\right)}\\
\gamma&=&\frac{2 \Gamma \left(\frac{2}{k+2}\right)^2}{\Gamma \left(\frac{1}{k+2}\right) \Gamma \left(\frac{3}{k+2}\right)}\ .
\eea
From (\ref{Uexpression}) we see that the different $p$ terms do not mix under crossing transformations. Imposing (\ref{Ucross}) for the $p=0$ term we have
\begin{multline}\label{crossingcondition}
A^+v_0^+(z)v_0^+(\zb) + A^-v_0^-(z)v_0^-(\zb) =\\
(A^+\beta^2+A^-\alpha^2)v_0^+(1-z)v_0^+(1-\zb) 
+ (A^+\gamma^2 +A^- \beta^2) v_0^-(1-z)v_0^-(1-\zb)\\
+(A^- \beta\alpha- A^+\beta \gamma)\left(v_0^+(1-z)v_0^-(1-\zb)+v_0^-(1-z)v_0^+(1-\zb)\right)
\end{multline}
giving the three equations
\bea
A^+\beta^2 + A^- \alpha^2 &=& A^+\\
A^+ \gamma^2 + A^- \beta^2 &=& A^-\\
A^- \alpha = A^+ \gamma \ .
\eea
From the structure of (\ref{v1mcross}) -- (\ref{v0pcross}) it is clear that the $p=1$ term gives the same equations. Finally we know that the $(-)$ channel is the vacuum. From this, it follows that $A^-=1$. We can thus solve for $A^+$ to get
\be
A^-= 1\ , \qquad A^+ = \frac{\Gamma \left(-\frac{2}{k+2}\right)^2 \Gamma \left(\frac{1}{k+2}\right) \Gamma \left(\frac{3}{k+2}\right)}{3 \Gamma \left(-\frac{3}{k+2}\right) \Gamma \left(-\frac{1}{k+2}\right) \Gamma \left(\frac{2}{k+2}\right)^2}\ .
\ee
Comparing to \cite{Zamolodchikov:1986bd}, we do indeed find that in their notation $A^-=N(1/2,1/2,1/2,1/2)h^-$ and $A^+=N(1/2,1/2,1/2,1/2) h^+$, which gives agreement.

\subsection{$U$ for $j_4=k/2$}\label{ss:Uk2}

For $j_4=k/2$, (\ref{Udiff}) simplifies to a single differential equation
\be\label{Uforj4}
(k+2)z(z-1)\frac{\partial}{\partial z}U_{p,\bar p} =
- (a_pz+b_p(z-1))U_{p,\bar p} 
\ee
where $a_p$ and $b_p$ are given by ($\ref{apbp}$).
The KZ equation thus reduces to a separable first-order differential equation. Equation (\ref{Uforj4}) and its $\bar z$ counterpart each have one basis solution. Together they give the general solution
\be
U_{p, \bar{p}}(z,\bar{z}) = A_p |1-z|^{- \frac{2 a_p}{k+2}} |z|^{- \frac{2 b_p}{k+2}}
\ee
Combining this with the general form \eqref{generalU} of the 4-point function, we obtain
\be
 U(x, \bar{x}; z, \bar{z}) =  A_p (x-z)^p (\bar{x}-\bar{z})^p |1-z|^{- \frac{2 a_p}{k+2}} |z|^{-\frac{2 b_p}{k+2}}
 \ee

Let us now fix $A_p$. 
We first rederive (\ref{Ajfrom3pt}). 
We note that the four-point function vanishes unless $j_1+j_2+j_3 \geq j_4$, hence $j_1+j_2+j_3-k/2\geq 0$ and $p_0 = j_1+j_2+j_3-k/2$. Hence we have $p=j_1+j_2+j_3-k/2$. The only channel is thus $j=k/2-j_3$.

The constant $A_p$ can then be expressed in terms of three-point functions. To see this, consider the leading term in the $z$ expansion of the four-point function
(\ref{CorrFromU}),
\be
A_j(x_{14}\xb_{14})^{2j_1}(x_{24}\xb_{24})^{-j_1+j_2-j_3+k/2}(x_{34}\xb_{34})^{-j_1-j_2+j_3+k/2}(x_{32}\xb_{32})^{j_1+j_2+j_3-k/2} (x \xb)^p
\ee
On the other hand we can use (3.17) in \cite{Zamolodchikov:1986bd} to expand $\Phi^{(3)}$ and $\Phi^{(4)}$. In that case, $R_0$ in (3.20) is trivial because $j+j_3-k/2=0$. Keeping only the $n=0$ term and combining (3.17) with (3.15), we find to leading order
\be
C(j_1,j_2,j)C(j,j_3,j_4)(x_{34}\xb_{34})^{j_3+j_4-j}(x_{12}\xb_{12})^{j_1+j_2-j}(x_{14}\xb_{14})^{j_1+j-j_2}(x_{24}\xb_{24})^{j_2+j-j_1}\ .
\ee
It is then straightforward to check that those two expressions indeed agree if we choose $A_j=C(j_1,j_2,j)C(j,j_3,j_4)$.

We can use this identification and crossing symmetry to fix the three-point functions.
We have
\be
 U(x, \bar{x}; z, \bar{z}) =  C(j_1,j_2,j)C(j,j_3,k/2) (x-z)^p (\bar{x}-\bar{z})^p |1-z|^{- \frac{2 a_p}{k+2}} |z|^{-\frac{2 b_p}{k+2}}
 \ee
where $b_p/(k+2) = \Delta_{j_1}+\Delta_{j_2}-\Delta_j$ and similar for $a_p$.

We first set $j_1=j_2$ and $j_3=j_4=k/2$. The three-point functions then become two-point functions because only the vacuum runs in the internal channel, giving
\be
 U(x, \bar{x}; z, \bar{z}) =   (x-z)^{2j_1} (\bar{x}-\bar{z})^{2j_1} |1-z|^{- 2j_1} |z|^{-4\Delta_1}
\ee
We then use (\ref{Ucross}) to obtain
\begin{multline}\label{U4231k12}
U_{4231}(x,z) = \frac{|1-z|^{4\Delta_1-4\Delta_4}}{|z|^{0}} ((1-x)(1-\xb))^{-2j_1+k}U_{1234}(1-x,1-z)\\
= (x-z)^{2j_1} (\bar{x}-\bar{z})^{2j_1}|z|^{-2j_1} |1-z|^{-4\Delta_4}\ .
\end{multline}
Using (\ref{GxzfromUxz}) we see that the internal channel has indeed dimension
\be
\Delta=\frac{(k/2-j_1)(k/2-j_1+1)}{k+2}\ .
\ee
Expanding (\ref{U4231k12}) in $z$ implies that $C(j,k/2,k/2-j)=1$. This agrees indeed with the three-point expression in \cite{Zamolodchikov:1986bd}.

\bibliographystyle{ytphys}
\bibliography{refmain}

\end{document}